\documentclass[aps,pra,a4paper,showpacs,twocolumn,10pt]{revtex4-1}
\usepackage[utf8x]{inputenc}    %
\usepackage{amssymb,amsmath}    %
\usepackage{graphicx}           %
\usepackage{textcomp}           %
\usepackage{color}              %
\usepackage{epstopdf}           %
\usepackage{bbm}                %
\usepackage{bm}			 %
\usepackage{mathtools}		 %
\usepackage{xfrac}

\definecolor{myurlcolor}{rgb}{0,0,0.7}
\definecolor{myrefcolor}{rgb}{0.8,0,0}
\usepackage[unicode=true,pdfusetitle, bookmarks=false,bookmarksnumbered=false,
bookmarksopen=false, breaklinks=false,pdfborder={0 0 0},backref=false,
colorlinks=true, linkcolor=myrefcolor,citecolor=myurlcolor,urlcolor=myurlcolor]
{hyperref}

\usepackage{lipsum}		%
\usepackage[caption=false]{subfig}		%

\usepackage{geometry}
\newcommand{\bra}[1]{\mathinner{\langle #1\rvert}}
\newcommand{\ket}[1]{\mathinner{\lvert#1\rangle}}
\newcommand{\Bra}[1]{\left< #1 \right|}
\newcommand{\Ket}[1]{\left| #1 \right>}
\newcommand{\ketbra}[2]{\mathinner{\lvert#1\rangle\langle #2\rvert}}
\newcommand{\Ketbra}[2]{\left|#1\middle>\middle<#2\right|}

\newcommand{\Sketbra}[3]{\Ket{#1}_{#3}\!\Bra{#2}}

\definecolor{orange}{RGB}{255,165,0}

\begin{document}
\title{Multipartite state generation in quantum networks with optimal scaling}
\author{J. Walln\"ofer, A. Pirker, M. Zwerger, and W.~D\"ur}
\affiliation{Institut f\"ur Theoretische Physik, Universit\"at Innsbruck, Technikerstr. 21a, A-6020 Innsbruck,  Austria}
\date{\today}

\begin{abstract}
We introduce a repeater scheme to efficiently distribute multipartite entangled states in a quantum network with optimal scaling. The scheme allows to generate graph states such as 2D and 3D cluster states of growing size or GHZ states over arbitrary distances, with a constant overhead per node/channel that is independent of the distance. The approach is genuine multipartite, and is based on the measurement-based implementation of multipartite hashing, an entanglement purification protocol that operates on a large ensemble together with local merging/connection of elementary building blocks. We analyze the performance of the scheme in a setting where local or global storage is limited, and compare it to bipartite and hybrid approaches that are based on the distribution of entangled pairs. We find that the multipartite approach offers a storage advantage, which results in higher efficiency and better performance in certain parameter regimes. We generalize our approach to arbitrary network topologies and different target graph states.
\end{abstract}
\pacs{} %
\maketitle

\section{Introduction \label{sec:intro}}

The distribution of entangled quantum states over large distances is a central task in quantum information processing. Initial studies have focused on point-to-point communication between a sender and a single receiver, where quantum repeaters \cite{Br98, ladd2006, Hartmann07,  Sa11, Azuma2015, Pirandola2016, Azuma2016, Knill96, Zw14, Muralidharan2014} have been developed to enable long-distance quantum communication over noisy channels in the presence of imperfect local control operations. Recent experimental developments, together with the promise of exciting applications of quantum technology, make large-scale networks or even a full-scale quantum internet a viable possibility. In such networks, not only the distribution of Bell states --which might be used for quantum communication between two parties via teleportation \cite{teleportation}, or for quantum key distribution \cite{e91, Lo99}-- is of relevance, but also the generation of multipartite entangled states shared among several nodes of the network. This opens the way for applications such as secret voting and secret sharing \cite{Hillery05}, conference key agreement \cite{Xu2014, Sun2016, Sun2016a}, clock synchronization \cite{Komar2014}, or distributed quantum computation \cite{Beals20120686}.

One important aspect in such quantum networks is efficiency - i.e. the required overhead for distributing entangled states shared between the communication partners. While a direct communication over noisy and lossy channels suffers from exponentially growing overheads with the distance (and hence small rates per channel usage), quantum repeater schemes (see e.g. \cite{Br98, ladd2006, Hartmann07,  Sa11, Azuma2015,Pirandola2016,Azuma2016}) or transmission of encoded information \cite{Knill96,Zw14,Muralidharan2014} allow for an efficient generation of bipartite entangled states between two communication partners with overheads that scale polynomially or polylogarithmically with the distance. 

Similar schemes have been proposed for the direct distribution of multipartite entangled states \cite{Kr06,epping_graph}, while in \cite{2drepeater} a quantum repeater scheme that is based on the usage of multipartite states that are connected and purified was introduced. While such schemes are in principle efficient, overheads significantly increase with the distance and rates are hence limited. 

In \cite{zwerger_big_data} a solution to this problem was proposed, where a quantum repeater scheme based on hashing --a deterministic entanglement purification protocol that operates on a large ensemble--- is used to establish long-distance quantum communication between a sender and a receiver with overheads per channel that do not grow with the distance. A key element in this approach is the measurement-based implementation of purification and connection processes \cite{zwerger_hashing}. 

Here, we introduce a similar scheme for the direct distribution of multipartite entangled states in a long-distance quantum network. Our scheme is capable of generating high-fidelity multipartite entangled target states shared between different nodes of the network with a constant overhead per channel, which is independent of the distance. To this aim, we combine the idea of so-called 2D quantum repeaters \cite{2drepeater} (or multipartite network repeaters) with the new type of repeaters based on hashing \cite{zwerger_big_data} in a multipartite setting. We analyze finite-size multipartite hashing schemes \cite{epptwocol, eppallgraphs, pirker_secure_hashing} which are a central element in our approach, and provide lower bounds on the global output fidelity of these entanglement purification protocols in terms of the number of input copies, initial fidelities and noise levels on the resource states of the measurement-based implementation. We illustrate the overall approach for the generation of 2D and 3D cluster states of growing size in the network, and also discuss the distribution of three-party GHZ states. However, the approach is not limited to regular networks, but can easily be adapted to other network topologies and different target states. Note, however, that the schemes we propose are not optimized for a near-term implementation with a very small number of resources, but require the storage and processing of a few hundred copies or more. In turn we obtain an efficient, scalable scheme that allows for the transmission of big quantum data in an intrinsically multipartite way.

We also discuss alternative schemes based on pairwise generation of entangled states, which are subsequently combined to form the desired multipartite target state (scheme B), and a hybrid approach that makes use of bipartite and multipartite elements (scheme C). We compare these three approaches, and develop optimized strategies to minimize the storage requirements of stations in the network. We also introduce and discuss variants of the multipartite protocol that use elementary building blocks of different size and shape (scheme A). The multipartite approach (A) offers an advantage over bipartite and hybrid approaches in this respect, as the storage requirements per repeater or network node are smaller. We analyze the performance of the different schemes with respect to reachable fidelities and obtainable number of output copies.

To this aim we concentrate on a scenario where the available quantum memory is limited. This is of practical relevance in a future quantum network. We consider the case where the overall storage capacity in the network is bounded, as well as a scenario where each node in the network has a quantum memory of the same (limited) size. In both cases, we identify parameter regimes (reachable target fidelity, channel noise and errors in resource states) where scheme (A) is superior to (B) and (C). We also remark that the hybrid approach (C) performs well in certain settings, and can even be more efficient than the multipartite approach. However, in particular in three-dimensional networks the storage advantage of the multipartite approach dominates.

The paper is organized as follows. In Sec. \ref{sec:concepts} we summarize basic concepts and required methods. We briefly review graph states and 2D repeater schemes, as well as the measurement-based implementation of repeater elements. We also discuss multipartite hashing schemes and analyze their performance for finite number of input states. In Sec. \ref{sec:problem_concepts} we define the problem setting and introduce different schemes to generate long-distance entangled states in the network using (A) an intrinsic multipartite approach, (B) a bipartite approach based on the distribution of entangled pairs and (C) a hybrid approach with bipartite and multipartite elements. In particular, we compare the different schemes in scenarios where the size of quantum memory is limited in Sec. \ref{sec:results}.  In Sec. \ref{sec:GHZ} we consider three-party GHZ states with growing distance. In Sec. \ref{sec:problem:blocks} we discuss explicit ways to efficiently obtain cluster states of growing size in networks with a 2D and 3D geometry and analyse them in a scenario where (i) the local memory per network node or (ii) the total memory is limited in Sec. \ref{sec:fromblocks} and \ref{sec:global_storage} respectively. In case of (ii) memory can be freely distributed among nodes to optimize performance. Furthermore, we discuss an additional scenario that starts from only Bell pairs distributed between neighboring parties in Sec. \ref{sec:frombell}. We provide a generalization of our approach to arbitrary network topologies and different kinds of target states in Sec. \ref{Sec:generalization}, and summarize and conclude in Sec \ref{sec:conclusion}.

\section{Concepts \label{sec:concepts}}

\subsection{Long-distance entangled states and quantum repeaters \label{sec:2drepeater}}
In this paper we focus on the generation of long-distance entangled states. For bipartite entangled states, such as Bell-pairs, this problem is addressed via quantum repeaters. Quantum repeaters were originally designed to establish Bell states over large distances in the presence of imperfections. There are various different architectures, in particular the approach based on quantum error correction \cite{Knill96} and the one based on iterative entanglement purification and swapping \cite{Br98}. In \cite{Hartmann07} and \cite{jiang_qec} the need for two-way communication in the original entanglement-based quantum repeater \cite{Br98} was removed. In all these approaches the local resources grow polylogarithmically or polynomially with the distance. 

The problem of creating graph states in quantum networks was adressed in \cite{epping_graph, cuquet_growth, meter_rec, pirkerdevices}. In particular, the work of \cite{epping_graph} shows how to generate long-distance graph states by generalizing the concept of quantum repeaters to multipartite entangled states. In \cite{cuquet_growth} different techniques were proposed for establishing small-scale graph states where clients need to merge small scale GHZ states for establishing the target graph states. Finally, \cite{meter_rec} proposes a recursive architecture for quantum repeater networks, with the aim of creating arbitrary graph states between its clients, while \cite{pirkerdevices} introduces a modular architecture that relies on quantum network devices to fullfil arbitrary graph state requests.

In \cite{2drepeater} a generalization of the 1998 quantum repeater protocol \cite{Br98} to multipartite states like GHZ or cluster states, was proposed. It is based on the preparation of elementary multipartite states, which are generated over short distances. Subsequently these states are purified via recurrence protocols. Then they are connected via measurements, similar to entanglement swapping, in order to obtain the desired multipartite state over a larger distance. This process is then iterated. The 2D quantum repeater shares the polynomial scaling of resources with the original quantum repeater scheme.
There is also a variant of the 2D quantum repeater, where one does not aim at preparing a multipartite state of fixed size (number of parties) but rather tries to let the number of parties grow with the distance \cite{2drepeater}. A major insight of \cite{2drepeater} was that there are parameter regimes for the noise in which a truly multipartite quantum repeater approach performs better than a bipartite approach (where one would first establish long-distance Bell pairs which are then used to create the desired multipartite state).

\subsection{Measurement-based implementation}
Measurement-based quantum computation \cite{briegel_meas, oneway, firstgraphstate} is a scheme for quantum computing which is based on adaptive single qubit measurements on a resource state. A prominent resource state for universal measurement-based quantum computing is the 2D cluster state \cite{oneway}. The read-in of an unknown input state can be achieved via Bell measurements, similar to teleportation \cite{teleportation}. Circuits which contain only Clifford gates and Pauli measurements can be implemented on resource states which contain only input and output qubits and no intermediate qubits. The map described by the circuit is performed solely by the Bell measurements at the read-in. Many quantum error correcting codes and entanglement purification protocols have this property.

\subsection{Quantum repeater based on hashing}

\begin{figure}
 \includegraphics[width=\columnwidth]{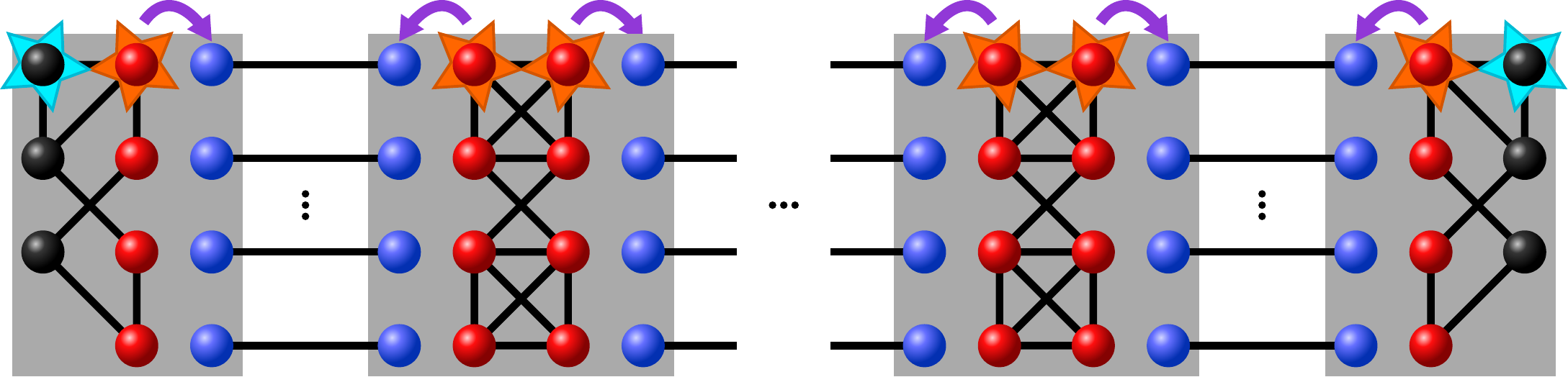}
 \caption{\label{fig:bip_mbqc_repeater} Illustration of the measurement-based implementation of the bipartite quantum repeater based on hashing. The initial Bell pairs (blue) are distributed to the neighboring stations and each station prepares a resource state with input (red) and output qubits (black). Note that local noise on the resource states (orange stars) can be shifted to the input states, but noise on the output qubits (blue stars) will still act on the output of the protocol.}
\end{figure}

In \cite{zwerger_big_data} a new quantum repeater scheme based on hashing \cite{bennett_hashing} with superior scaling was introduced. The hashing protocol for bipartite entanglement purification operates on a large number of Bell pairs and outputs $m = n(1-S)$ perfect Bell pairs in the asymptotic limit. Here, $n$ denotes the number of input pairs and $S$ their entropy. The protocol has a non-zero yield (ratio of output and input pairs in the asymptotic limit) in contrast to recurrence protocols \cite{bbpssw, dejmps}, where the yield vanishes.
In the novel quantum repeater protocol, the nested levels of entanglement purification and swapping are replaced by a single, combined step of entanglement purification via the hashing protocol \cite{bennett_hashing} and simultaneous swapping of all pairs. This setup is depicted in Fig. \ref{fig:bip_mbqc_repeater}. In this way the scaling of the local resources per transmitted qubit is reduced from polynomial to constant. 
We remark that there are no nested repeater levels in this scheme as the whole scheme is implemented in a single step.

\subsection{Graph states and the graph state basis \label{sec:graphs}}
Graph states are a special subset of quantum states that are associated with mathematical graphs. 
A graph $G = (V,E)$ consists of $N$ vertices $V$ and edges $E$ and the corresponding graph state $\Ket{G}$ is given by:
\begin{equation}
 \Ket{G} = \prod_{ \left\{a,b \right\} \in E } U_{\textrm{CZ}}^{ab} \Ket{+}^{\otimes N}
\end{equation}
with $U_{\textrm{CZ}}^{ab}=\ketbra{0}{0}^{(a)} \otimes \mathbbm{1}^{(b)} + \ketbra{1}{1}^{(a)} \otimes Z^{(b)}$, the controlled phase gate acting on qubits $a$ and $b$, and $\Ket{+} = 1/\sqrt{2} (\Ket{0} + \Ket{1})$.
We call a graph state $k$-colorable if $k$ is the smallest number of colors needed to color each vertex in its associated graph such that no two vertices of the same color are connected by an edge.

The \textit{graph state basis} is an orthonormal basis that is defined with respect to a certain graph $G$:
  \begin{equation}
  \Ket{\bm{\mu}}_G = \prod_{j \in V} \left( Z^{(j)} \right)^{\mu_j} \Ket{G}
  \label{eqn:graphstatebasis}
 \end{equation}
where $\bm{\mu}=(\mu_1, \mu_2, \dots, \mu_N) \in \{0,1\}^N$. 

For the purpose of this work we concern ourselves with states diagonal in the graph state basis, which we write as:
\begin{equation}
  \rho = \sum_{\bm{\mu}} \lambda_{\bm{\mu}} \Sketbra{\bm{\mu}}{\bm{\mu}}{G}
\end{equation}
This diagonal form can always be achieved by depolarization.

\subsubsection{Connecting graph states \label{sec:graphconnect}}
In this work we will often mention \textit{connection operations} to connect graph states. We employ two different connection operations, one where two qubits are merged into one, and one where both connection qubits are projected out.

For the first method one takes two qubits $a$ and $b$, which correspond to vertices of the graph states that we want to connect, and applies the CNOT operation $\textrm{CNOT}^{a \rightarrow b} = \ketbra{0}{0}^{(a)} \otimes \mathbbm{1}^{(b)} + \ketbra{1}{1}^{(a)} \otimes X^{(b)}$ with qubit $a$ as the source and qubit $b$ as the target, followed by a $Z$-measurement on qubit $b$. The resulting state 
is a graph state (up to local Clifford corrections) that corresponds to a graph without vertex $b$ and the neighborhood $N^\prime_a$ (i.e. all vertices connected to $a$ with an edge) of vertex $a$ is now given by $N_a \bigcup N_b - N_a \bigcap N_b$.

The second kind of connection operation works in a similar fashion. But first one needs to transform the initial graph states by applying the local complementation operation $\tau$ (see e.g. \cite{He06}) on both qubits $a$ and $b$, which are initially not connected by an edge. Then, just as before, one applies $\textrm{CNOT}^{a \rightarrow b}$ followed by measuring qubit $b$ in the $Z$-basis. Finally, one measures qubit $a$ in the $Y$ basis. The resulting state is, again, a graph state that corresonds (up to local Clifford corrections) to a graph with vertices $a$ and $b$ removed and changed edges according to: $N^\prime_i = N_i \bigcup N_b - N_i \bigcap N_b$ for all $i$ that were initially in $N_a$ and $N^\prime_j = N_j \bigcup N_a - N_j \bigcap N_a$ for all $j$ in $N_b$.
In this work we only use this second connection operation in the context of GHZ states as depicted in Fig. \ref{fig:whole_setup},  where its effect can be understood as mapping two $n$-qubit GHZ states to a state that is local Clifford-equivalent to a GHZ state with $(2n - 2)$ qubits. 

See \cite{He06} for a detailed summary on the properties of graph states and how noise and other transformations acting on graph states can be described. 

\subsection{Noise model}
We model noise via local depolarizing noise (LDN), which is given by the map ${\cal{D}}$ and acts on a qubit with density matrix $\rho$ in the following way:
\begin{equation}
{\cal{D}}(q)\rho= q\rho + \frac{1-q}{4} \left(\rho+X\rho X + Y\rho Y + Z\rho Z \right).
\label{eqn:ldn_def}
\end{equation}
Here the error parameter $q \in [0, 1]$ describes the strength of the noise, $q=1$ corresponds to no noise and $q=0$ to complete depolarization and $X, Y, Z$ refer to the usual Pauli matrices. It should be noted that LDN can be interpreted as a worst case estimate for local noise \cite{effectivenoise}.

We describe the noise that occured during the initial distribution of the state, for example by sending them through noisy quantum channels, by LDN with error parameter $q$ acting on all qubits independently, e.g. a noisy graph state is described by
\begin{equation}
\prod_{i=1}^{n} {\cal{D}}_i(q)\ket{G}\bra{G},
\end{equation}
where the subindex refers to the qubit on which ${\cal{D}}_i$ acts on. 

The second source of noise we consider is the imperfections of the resource states we use for the measurement based implementation of our schemes. We describe the noise on the resource states, which are generated locally at each party, by LDN with a different error parameter $p$ acting on all qubits of the resource state.

An important property of LDN is that its location can be exchanged if it is followed by a Bell projection \cite{Zw13}, i.e.,
\begin{equation}
P_{1,2}{\cal{D}}_1(q) \rho = P_{1,2}{\cal{D}}_2(q) \rho,
\end{equation}
where $\rho$ is some density matrix and $P_{1,2}$ denotes a projector on one of the four Bell states, acting on qubits $1$ and $2$. The noise on the input qubits of the resource states can thus be (formally) shifted to the input states instead (see Fig. \ref{fig:bip_mbqc_repeater}). 

This allows one to interprete a noisy, measurement-based implementation of an entanglement purification protocol or a quantum repeater as additional noise acting on the input states followed by the perfect protocol and noise on the output qubits. This is due to the fact that the resource states for such protocols have only input and output qubits (see above) and the processing of the input states is performed via Bell measurements. For more details see \cite{Zw13}.

\subsection{Multipartite quantum hashing protocol \label{sec:hashing}}

Quantum hashing protocols \cite{bennett_hashing,epptwocol,eppallgraphs,chen_hashing,maneva_hashing,hostens_hashing,glancy_hashing,hostens_breeding} are a type of entanglement purification protocol. They are based on the quantum analogon of the noiseless coding theorem \cite{noiseless_coding, quantum_coding} and rely on the fact that it is exponentially likely that the input states are in a so-called \textit{likely subspace}. One notable feature of the hashing protocols is that, unlike recursive entanglement purification approaches (e.g. \cite{bbpssw, dejmps}), they can deterministically provide a non-zero asymptotic yield. Unfortunately due to the nature of operations required, a gate-based implementation cannot tolerate any imperfections in the operations. However, recently it was shown that a measurement-based implementation makes hashing protocols practical in the presence of imperfections \cite{zwerger_hashing}.

The multipartite quantum hashing protocol that we utilize in this work is described in \cite{epptwocol,eppallgraphs} and works for graph states. The most prominent difference to the bipartite case is that to purify a $k$-colorable graph state one separate subprotocol for each color is necessary to obtain all the necessary information \cite{eppallgraphs}.

We will briefly describe the basic mechanisms of the bipartite hashing protocol and discuss how the multipartite hashing protocol differs from the bipartite case and what challenges arise from them. See Appendix \ref{sec:multidetails} for additional details on estimating fidelities.

The bipartite protocol that we consider here \cite{bennett_hashing} roughly works as follows: Starting from $n$ input copies of a noisy Bell state $\rho$ with sufficiently high fidelity with respect to the desired Bell state $\Ket{\Phi^+}$, we want to extract $m$ copies of the perfect $\Ket{\Phi^+}$ state. First, the possible Bell states are encoded as $a_i = 00,01,10,11$ for $\Ket{\Phi^+}, \Ket{\Psi^+}, \Ket{\Phi^-}, \Ket{\Psi^-}$ respectively and the coefficients of $\rho^{\otimes n}$ can be interpreted as probabilities of being in a state corresponding to a bitstring $\widetilde{a}=a_1a_2\dots a_n$. Now one proceeds to measure random subset parities to determine the string $\widetilde{a}$ by applying bilateral CNOT operations and measuring out some of the copies. This approach works because one does not have to consider all possible strings $\widetilde{a}$. Since it is exponentially likely that $\widetilde{a}$ will lie in the likely subspace, one only needs to differentiate between those strings. It is possible to obtain $m = n (1 - S(\rho) - 2 \delta )$ output copies this way, where $S(\rho)$ is the von Neumann entropy of $\rho$ and $\delta > 0$ is allowed to approach $0$ as $n$ tends to infinity. For $n \rightarrow \infty$ one can deterministically obtain a yield of $Y = m / n = 1 - S(\rho)$. For finite $n$ however, there is a chance that the bitstring $\widetilde{a}$ either falls outside of the likely subspace or cannot be successfully distinguished from other bitstrings in the likely subspace. This case is especially relevant for this work as we want to consider scenarios with limited storage capacities, i.e. small $n$. One can find a lower bound for the success probability \cite{pirker_secure_hashing, zwerger_big_data}, which leads directly to a lower bound $f(a, n, \delta)$ for the \textit{global fidelity}, i.e. the overlap of the output pairs with $\Ket{\Phi^+}^{\otimes m}$. While $\delta$ can be chosen freely in this context, it is important to consider that $\delta$ is directly connected to the number of output pairs $m$.

For the multipartite protocol \cite{epptwocol, eppallgraphs} one can enumerate the states in the graph state basis (see \eqref{eqn:graphstatebasis}) in a similar way as with the different Bell states previously. However, the main limitation one faces is that arbitrary substring-parities cannot be extracted via local measurements, which is why it is necessary to consider one separate bitstring for each vertex in the graph, e.g. $a^{(k)} = a_1^{(k)} \dots a_n^{(k)}$ for the $k$-th vertex. Information about all bitstrings corresponding to one color can be extracted simultaneously, which is why one subprotocol per color is sufficient. The relevant entropy for $a^{(k)}$ is the entropy of the $k$-th bit in the graph state basis vector:
\begin{equation}
 S_k = S(a^{(k)}) = - \lambda_{k,0} \log_2 \lambda_{k,0} - \lambda_{k,1} \log_2 \lambda_{k,1}
\end{equation}
where $\lambda_{k,i} = \sum_{\mu_{j \neq k}} \lambda_{\mu_1 \dots \mu_{k-1} i \mu_{k+1} \dots \mu_N}$ is the probability that the $k$-th bit in the graph state basis vector $\bm{\mu}$ equals $i$. 
The whole protocol can only be considered successful if all the separate bitstrings are identified correctly, therefore we obtain a lower bound for the overall fidelity $\prod_k f(a^{(k)}, \delta_k)$.
For two-colorable graph states the asymptotic yield is given by $Y = \lim_{n\rightarrow \infty} m/n = 1 - S_A - S_B$, where $S_A = \max_{k \in A} S_k$ and $S_B = \max_{k \in B} S_k$ for vertices in colors $A$ and $B$, respectively.

\section{Schemes to establish multipartite states in a quantum network  \label{sec:problem_concepts}}

We consider a quantum network, i.e. a set of spatially separated parties that are connected via quantum channels. The parties are equipped with storage devices (quantum memory), and are capable to manipulate their stored states. We assume that the transmission between parties takes place via a noisy quantum channel with error parameter $q$, while the local manipulation is done in a measurement-based way. There, we assume noisy resource states with local noise per qubit specified by error parameter $p$. The goal is to establish multipartite entangled states, specifically some graph states, shared between all (or some of the) parties.
We will consider this problem in different settings: First, when storage is unlimited and we are interested in scaling of resources and overhead only. Second, in a scenario where storage is limited (either in total, or at each station).

The schemes we propose here are capable to establish multiple copies of graph states over arbitrary distance, with an optimal scaling. The overhead per transmitted state is only constant. In this sense, our schemes are a generalization of the 1D repeater scheme of Ref. \cite{zwerger_big_data} that allows one to establish Bell pairs between two distant parties to arbitrary target states and network geometries. The key element of our scheme is the generation of elementary building blocks, which are then purified via hashing, and merged or connected. In contrast to recurrence-based repeater schemes, no nesting or repeater levels ---which lead to polynomial or polylogarithmical scaling of resources with the distance--- are required. The fast convergence of the hashing protocol to maximally entangled states allows one to connect or merge all elementary building blocks in a single step, and obtain multiple copies with an overhead per transmitted qubit that is constant. %

These elementary building blocks could be Bell pairs shared between neighboring parties, or also multi-party states such as GHZ or cluster states. The crucial observation here is that the aforementioned properties of the hashing purification protocol for Bell-pairs, i.e. finite yield and exponentially fast convergence towards unit fidelity, also hold for the multipartite hashing protocol. Hence these states can be merged or connected in an arbitrary way. 
In the following we describe the three schemes for purifiying multipartite graph states as outlined in Sec. \ref{sec:intro}. All schemes have in common that their ultimate goal is to obtain a fixed multipartite entangled state after their completion, for example a 3-party GHZ state shared between distant parties, or a cluster state shared between all parties of the network. The schemes differ in the elementary building blocks, i.e. whether they work with multipartite entangled states (scheme A), or with bipartite entangled states (schemes B and C). In addition, the purely bipartite scheme (B) performs purification and merging/connection in two separate steps, while in schemes A and C these two steps are performed simultaneously in a measurement-based implementation. This is in fact crucial to obtain a scalable scheme. At intermediate stations the perfect hashing plus connection/merging is performed on a slightly noisier input state, and no additional errors are introduced as there are no output particles. In contrast, when performing hashing and connection/merging in two steps, additional noise on output states is introduced, and leads to a (exponentially) vanishing fidelity when combining multiple elementary building blocks. The schemes also differ in their storage requirements, as storing a multipartite state requires less memory than the storage of Bell pairs shared between different parties. 

\subsection{Scheme A \label{sec:problem_concepts:a}}
\begin{figure*}
 \centering
 \subfloat[\centering \label{fig:scheme_a} Scheme A]{\includegraphics[width=0.26\linewidth]{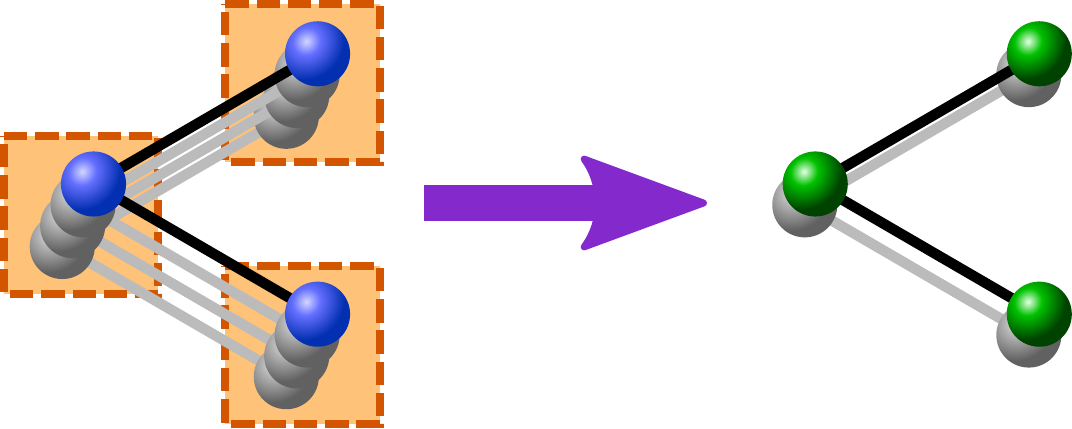}}
 \hfill
 \subfloat[\centering \label{fig:scheme_b} Scheme B]{\includegraphics[width=0.36\linewidth]{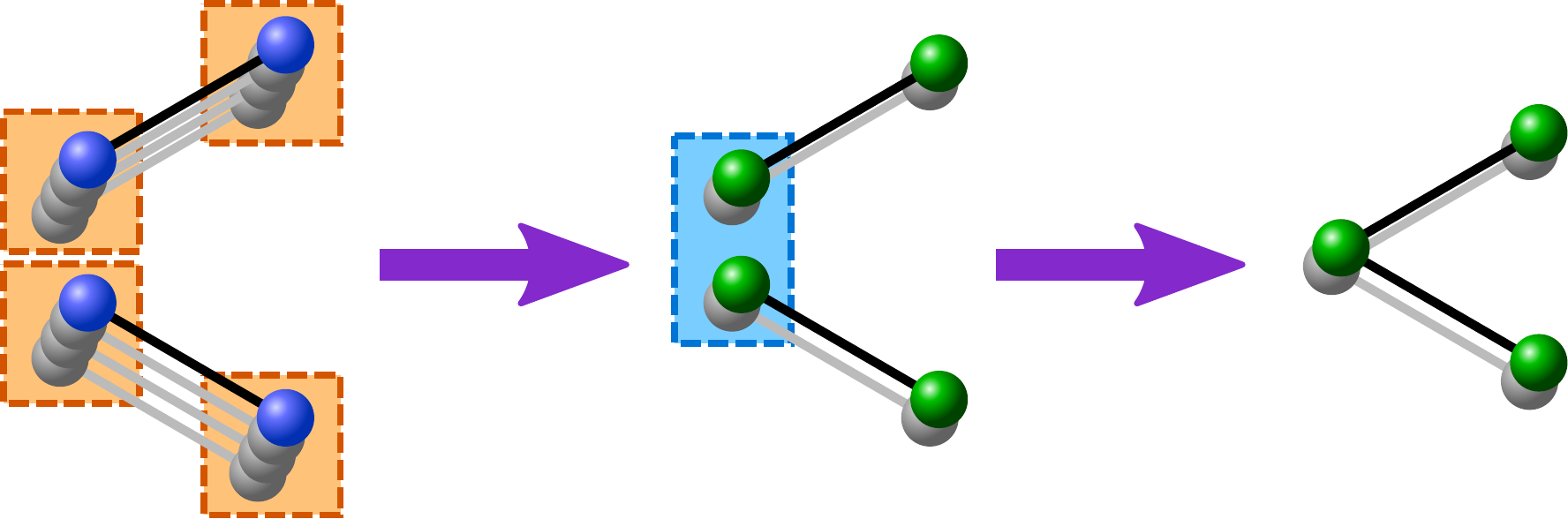}}
 \hfill
 \subfloat[\centering \label{fig:scheme_c} Scheme C]{\includegraphics[width=0.26\linewidth]{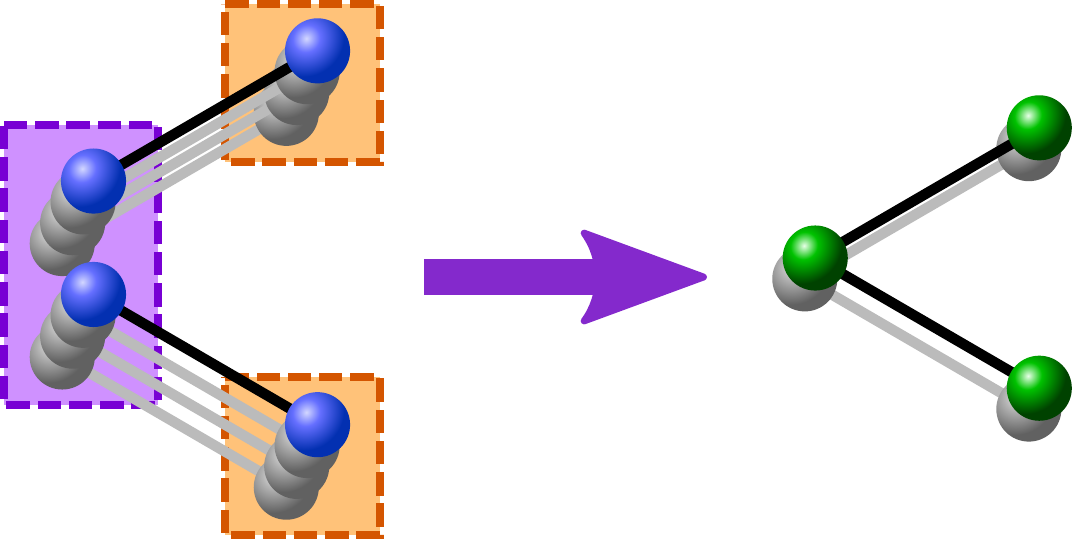}}
 \caption{The figure depicts the different schemes we consider. (a) Starting with multiple copies of a multipartite state (GHZ states in this example) each station implements the multipartite hashing protocol (marked by the orange box) in a measurement-based way to obtain fewer copies of that state with higher fidelity. (b) Before the protocol neighboring stations share several copies of Bell pairs. The stations implement the bipartite hashing protocol (orange boxes) in a measurement-based way to obtain purified Bell pairs. Then, in a separate step, the stations prepare another resource state for merging the Bell-pairs (blue box) into the desired multipartite state. (c) The neighboring stations share, like in scheme B, several copies of noisy Bell pairs. Each station performs the measurement based implementation of the multipartite hashing protocol (orange boxes) or the hashing protocol combined with the merging operation (purple box) where connections are needed.}
\end{figure*}

In scheme A the stations of the network share several copies of a noisy multipartite entangled state with neighbouring stations. Furthermore, each station prepares a resource state for the measurement-based implementation of the hashing protocol for two-colorable graph states (end nodes), or hashing followed by state merging or state connection at intermediate nodes. Each station of the network now couples the copies of the multipartite entangled state to the resource state via Bell-measurements, thereby purifying the states (see Fig. \ref{fig:scheme_a}), and merging or connecting the output states. As long as the initial fidelity (and fidelity of resource states) is sufficiently large, the protocol deterministically generates several copies of multipartite entangled target states.

\subsection{Scheme B \label{sec:problem_concepts:b}}

In contrast to scheme A all neighbouring stations share impure Bell-pairs in scheme B. This scheme comprises the following steps: First, each station prepares several copies (one for each neighbouring station) of the resource state for the measurement-based implementation of the hashing protocol for Bell-pairs. Next, each station couples the resource states to the impure Bell-pairs via Bell-measurements, thereby performing the hashing protocol. Assuming the initial fidelity is sufficiently high, one obtains fewer, but purified copies of Bell-pairs. Finally, depending on the multipartite target state, the stations merge or connect the purified Bell-pairs to obtain copies of the multipartite state (see Fig. \ref{fig:scheme_b}). This is done in a separate step that requires an additional resource state for the measurement-based connection of the Bell-pairs.

Observe that noise will act in this scenario at two steps: noise from the hashing protocol, and noise from the merging operation. In addition, one needs to store qubits corresponding to Bell states shared between all neighboring parties instead, which leads to a larger storage requirement for this scheme. %

\subsection{Scheme C \label{sec:problem_concepts:c}}
Finally we discuss scheme C, which is a combination of the bipartite and multipartite approach. In particular, we can obtain a combined resource state for both the bipartite hashing protocol and the merging or connection operation applied afterwards. This state can be readily obtained by virtually combining the two resource states via Bell-measurements. This leads to a smaller resource state which performs both tasks within a single step (see Fig. \ref{fig:scheme_c}). This has the advantage that noise from imperfect resource states will act only once, in contrast to scheme B where noise will act on the output of the purification protocol twice. Note that the architecture has to be flexible enough to generate different resource states depending on the desired graph state to profit from this advantage in general. The scheme however works with initial Bell pairs as input states, and hence has the same memory disadvantage as scheme B when compared to the truly multipartite scheme A. 

\subsection{Scaling of schemes}

We now discuss the scaling of the local resources of the different schemes. We are interested in the scaling with the distance of the number of qubits which need to be stored/processed at each repeater station in order to obtain a target state with a fidelity exceeding a fixed value. For the fidelity we choose the global, private fidelity $F_{\rm{gp}}$, which is the fidelity of the ensemble of output target states prior to the action of the noise on the output qubits of the resource states relative to the desired tensor product of target states (for more details see \cite{zwerger_big_data, pirker_secure_hashing}).

From this one can already see that scheme B is not scalable because for any non-zero value of noise one obtains Bell pairs with non-unit fidelity. These pairs are then further processed and the fidelity will drop similar to the case of swapping imperfect Bell pairs. The situation is different for scheme A and C, where the purification of elementary states and their further processing (merging) are executed in a single, simultaneous step. Here, a lower bound on the global, private fidelity $F_{\rm{gp}}$ can be obtained from the probability that all hashing protocols in the entire quantum repeater protocol succeed simultaneously. For a quantum repeater where $N$ states are merged one obtains
\begin{equation}
F_{\rm{gp}} \ge (1 - \alpha \text{exp}(-\beta n \delta^2 ))^N \approx 1 - N \alpha \text{exp}(-\beta n \delta^2)
\end{equation}
for a sufficiently large number $n$. Here $n$ is the number of input states of the hashing protocol and $\alpha$ and $\beta$ depend on properties of the input states and certain choices within the hashing protocol. For more details see Appendix \ref{sec:multidetails}. One can ensure that $F_{\rm{gp}}$ is close to one, i.e., $F_{\rm{gp}} \ge 1- \epsilon$ by choosing $n$ such that $N \alpha \text{exp}(-\beta n^{1/2}) < \epsilon$. This shows that one has to choose the number of input states according to $N$ (which is usually related to the distance) and the desired fidelity. For larger values of $N$ one will require (logarithmically) larger values of $n$. Note, however, that this will also lead to a larger number of output states $m$. Thus the overhead, i.e. the ratio $\tfrac{n}{m}$, becomes constant for large $n$, which is the optimal scaling.

We emphasize that this result is in contrast to previous schemes based on recurrence protocols that have polynomially or polylogarithmically scaling overheads.

\section{Application to a limited storage scenario \label{sec:results}}

Here, we address a setting with memory restrictions, which are crucial to consider for practical implementations.
In particular, we consider a situation where the memory sizes of the intermediate repeater stations are limited. This implies that an efficient strategy for memory usage and consumption needs to be applied to obtain the target state with highest possible fidelity. 
Due to the nature of the hashing protocols, the estimated fidelities with a limited amount of input copies are only bounds (see Appendix \ref{sec:multidetails}). \newline
In Sec. \ref{sec:GHZ} we investigate the application of the multipartite hashing protocol to GHZ states and discuss the distribution of a 3-qubit GHZ state over long distances. Then, in Sec. \ref{sec:cluster}, we investigate multiple scenarios of generating two-dimensional and three-dimensional cluster states from smaller building blocks. 
Since the memory usage is of utmost importance, we identify building blocks for cluster states that need to store as few qubits as possible in section \ref{sec:problem:blocks}. 
\subsection{3-qubit GHZ state \label{sec:GHZ}}
To begin the comparison between multipartite and bipartite approaches, we investigate the application of the multipartite hashing protocol to the 3-qubit GHZ state. The GHZ state is a truly multipartite entangled state and furthermore it is local-Clifford equivalent to a two-colorable graph state (see Fig. \ref{fig:scheme_a}), which allows the direct application of the hashing protocol for graph states. The GHZ state makes for an interesting graph state to analyze not only because of its simplicity but it is also at the core of the GHZ based two-dimensional repeater scheme of \cite{2drepeater}.
First, we take a look at input states for which each qubit has been affected by local depolarizing noise with parameter $q$. This corresponds e.g. to a situation where a perfect GHZ or Bell pair is generated locally by some source, and the states are distributed via noisy channels to the parties that are involved in the protocol.

\begin{figure}
 \includegraphics[width=\columnwidth]{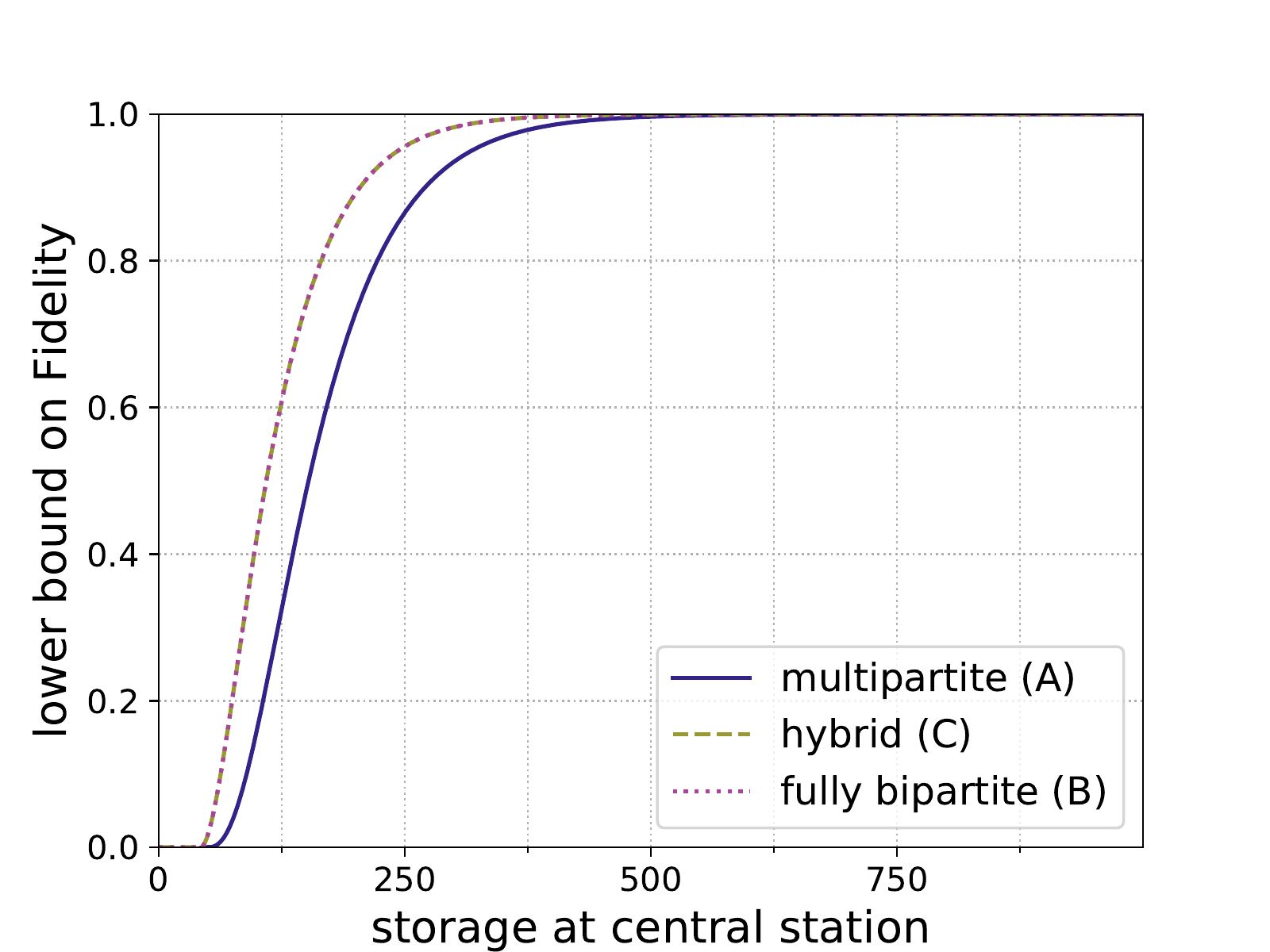}
 \caption{\label{fig:ghz_white_noise} Reachable fidelities for a 3-qubit GHZ state for bipartite and multipartite $n \rightarrow 1$ hashing protocols with $q=0.98$ local depolarizing noise per qubit on the input states.}
\end{figure}

Even in this simple model the storage advantage of employing a multipartite approach becomes apparent. When relying only on Bell pairs (scheme B), the station, where the Bell pairs are connected to obtain the final GHZ state, needs to store twice as many qubits. This means that for each separate bipartite hashing protocol between the different parties only half as many copies (when compared to scheme A) of Bell states will be available if the storage capacity of the stations is limited.
However, for this particular case the storage advantage is not sufficient for the multipartite approach to obtain better fidelities when using the hashing protocol as a $n \rightarrow 1$ protocol, which is done by choosing $\delta_A$ and $\delta_B$ such that the number of output copies $m=1$. \footnote{The advantage of looking at the $n \rightarrow 1$ variant is that the output fidelity directly corresponds to the state fidelity. However, the usual use case of the quantum hashing protocol is when multiple output copies are required.}. These results are depicted in Fig. \ref{fig:ghz_white_noise} for a fixed choice of $q=0.98$.

\begin{figure}
 \includegraphics[width=\columnwidth]{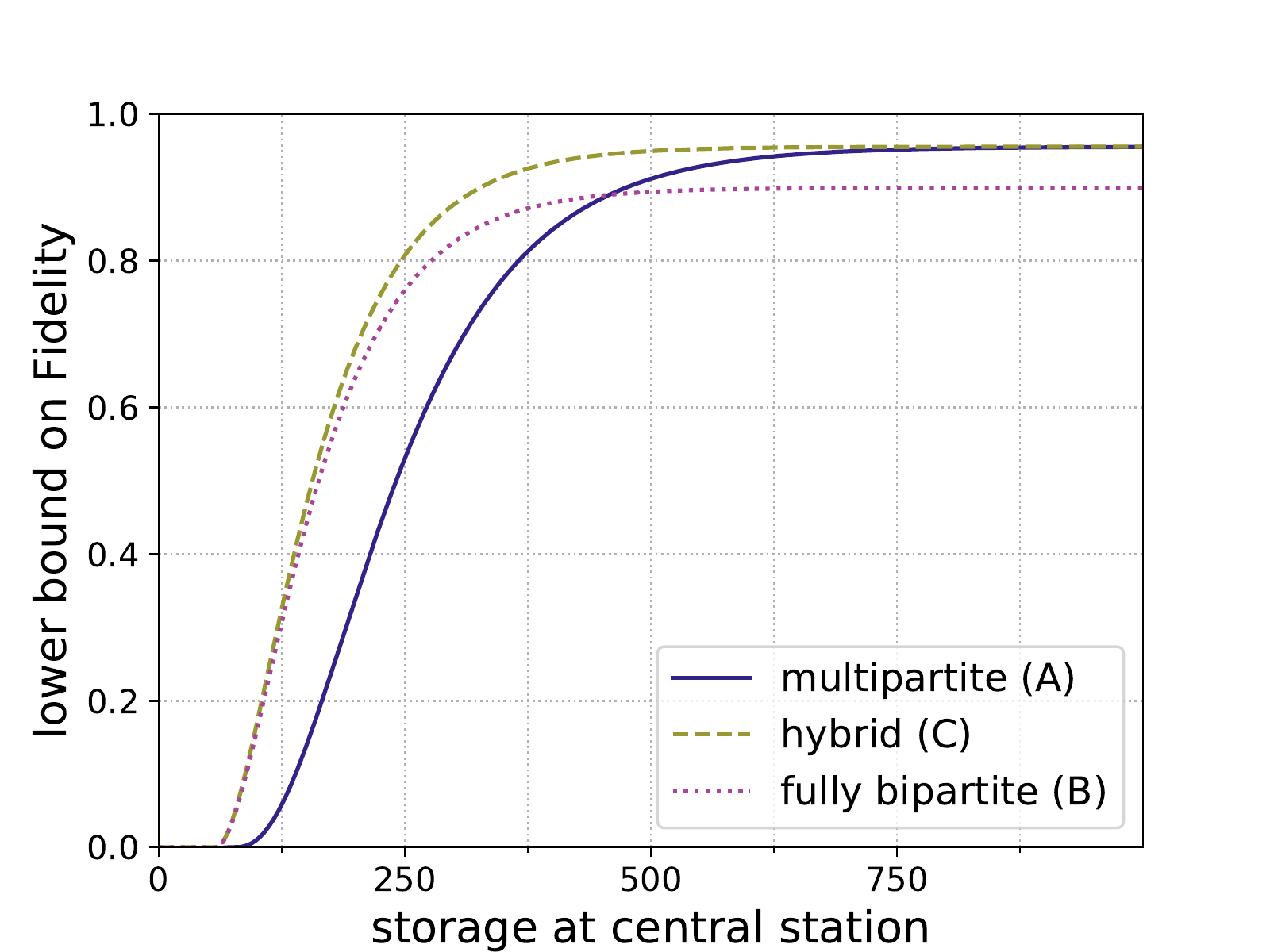}
 \caption{\label{fig:ghz_imperfect_resource} Reachable fidelities for a 3-qubit GHZ state with imperfect resource states for $n\rightarrow1$ hashing protocols. The initial states are affected by local depolarizing noise with error parameter $q=0.99$ and the resource state with error parameter $p=0.98$.}
\end{figure}

When considering imperfect resource states, we can see an interesting development depending on whether the bipartite approach is given the information of what the target state is before building the resource states (scheme C), see Fig. \ref{fig:ghz_imperfect_resource}. If the bipartite approach is not able to adapt to different output states (scheme B), the states will need to be connected afterwards, which exposes them to additional noise. Therefore, the qubits at the station connected to the two other parties will be affected by additional noise after the purification has taken place. If this step is also implemented in a measurement-based way, the additional noise takes the form of the imperfections in the extra resource state that needs to be created to perform the connection. In Fig. \ref{fig:ghz_imperfect_resource} it is obvious that the reachable fidelity of such a bipartite approach that is oblivious to the final use of their purified states has a smaller reachable fidelity than the multipartite approach. If the bipartite approach is allowed to use custom resource states that implement the bipartite hashing protocol as well as the connection operation, which we called scheme C, then the bipartite approach is superior for this case.

\subsubsection{Restricted error model \label{sec:restricted_error}}

We consider a restricted error model for the graph state version of the GHZ state and $n\rightarrow1$ protocols. The main reason why the multipartite approach performs worse is that a separate protocol is needed for each of the two colors, which the small storage advantage in this scenario is not able to overcome. Now, we look at a situation where the noise only affects one color, namely only $Z$-noise acting on the outer two qubits (the qubits at the dangling ends on the right in Fig. \ref{fig:scheme_a}). This situation could e.g. arise when distributing two qubits of a locally generated GHZ state via a noisy channel where dephasing is predominant, and the channel can be described by a phase-flip channel.

\begin{figure}
 \includegraphics[width=\columnwidth]{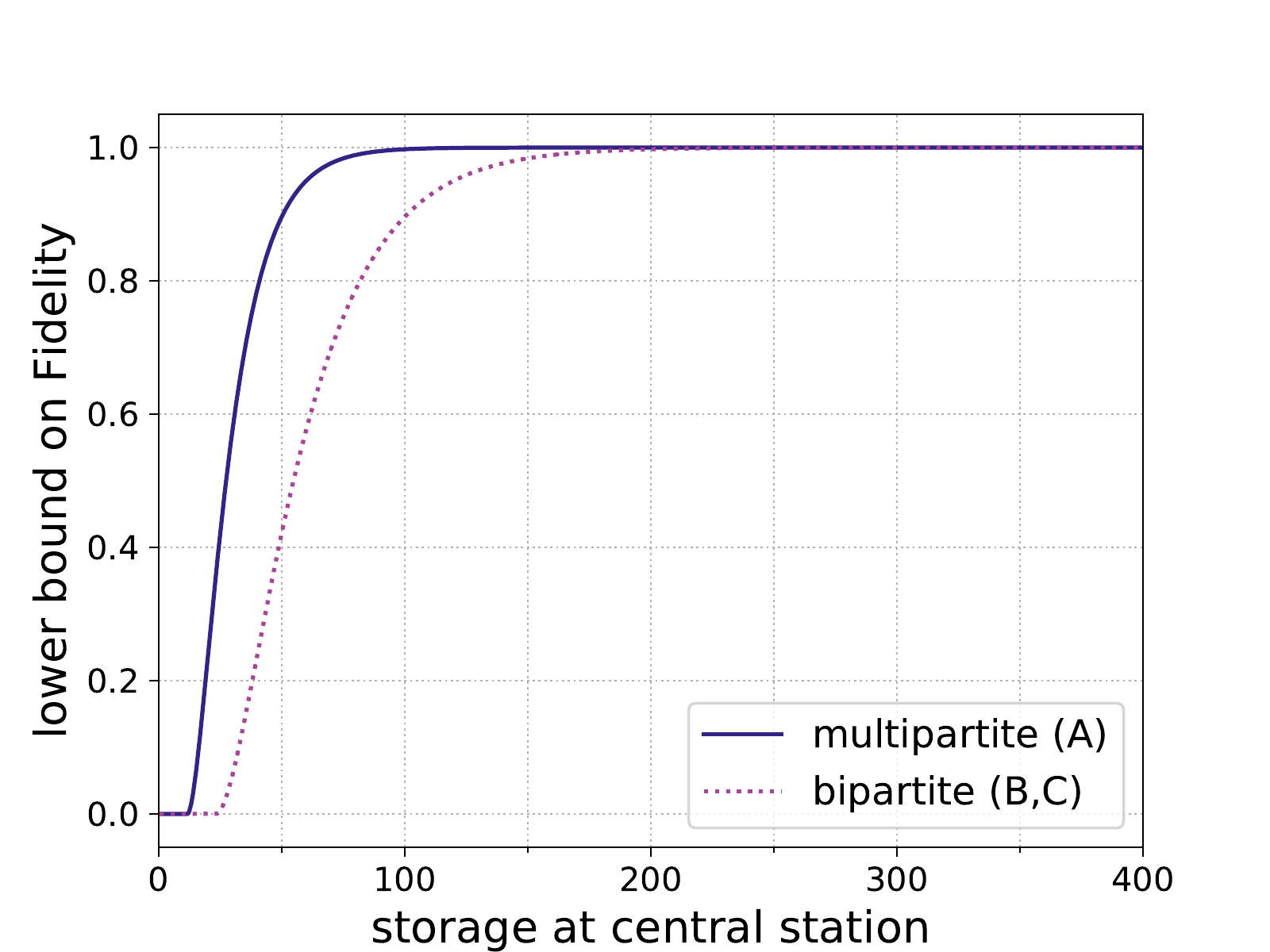}
 \caption{\label{fig:ghz_only_z} Reachable fidelity of a 3-qubit GHZ state for $n\rightarrow1$ hashing protocols, where the initial states are only effected by $Z$-noise with error parameter $q=0.98$ on the outer qubits. This avoids the requirement to use a second subprotocol for the multipartite protocol.}
\end{figure}

In this case the second protocol is not needed and the storage advantage immediately translates to higher reachable fidelities as shown in Fig. \ref{fig:ghz_only_z}.

However, if there is an additional, small amount of $X$-noise, also only acting on the outer qubits, it becomes mandatory to use the second subprotocol. We describe the action of the noisy channel by:

\begin{equation}
 \mathcal{E} \rho = (1-0.02001) \rho + 10^{-5} X \rho X + 0.02 Z \rho Z
\end{equation}

\begin{figure}
 \includegraphics[width=\columnwidth]{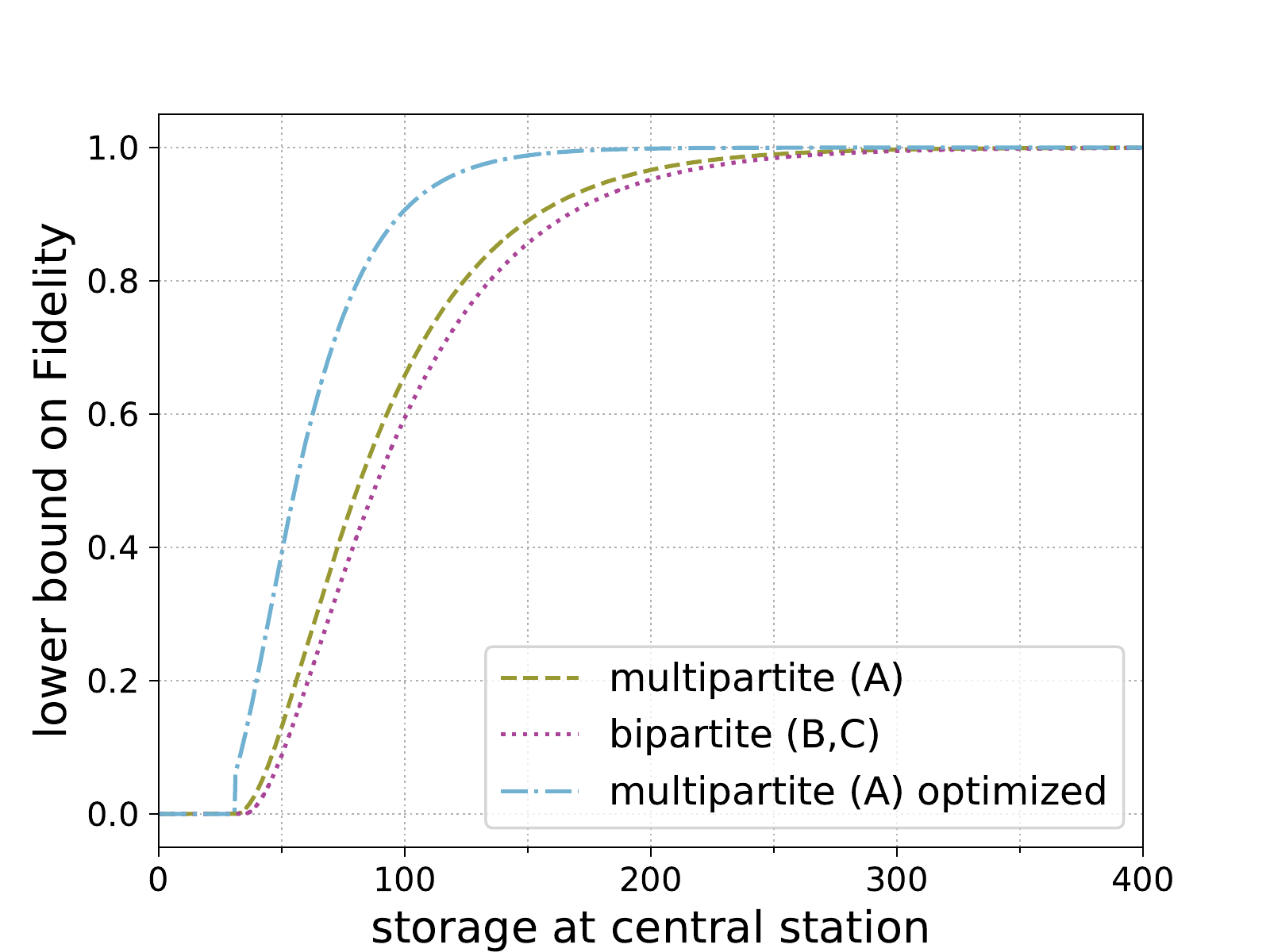}
 \caption{\label{fig:ghz_optimized_deltas} Reachable fidelity of a 3-qubit GHZ state, where the outer qubits of the intial states are effected by very biased noise with $2\%$ $Z$-noise and $10^{-3}\%$ $X$-noise . For very asymmetric noise optimizing how the additional input copies are distributed can significantly improve performance.}
\end{figure}

In this case with very asymmetric noise, it is important to distribute the additional input states that we have available in a $n\rightarrow1$ protocol appropriately among the subprotocols, i.e. choosing the same $\delta$ for both subprotocols (i.e. $\delta_A = \delta_B$) is a bad choice in this case. Fig. \ref{fig:ghz_optimized_deltas} illustrates that in such a situation some of the advantage of the multipartite approach (A) still remains. It highlights in particular that much can be gained by optimizing the distribution of the leftover copies, i.e. the choice of $\delta_A$ and $\delta_B$ for the different subprotocols. For local depolarizing noise, which is symmetric, the improvement gained by the optimization is negligible.

\subsubsection{Long-distance GHZ on triangular network}
\begin{figure}
 \includegraphics[width=\columnwidth]{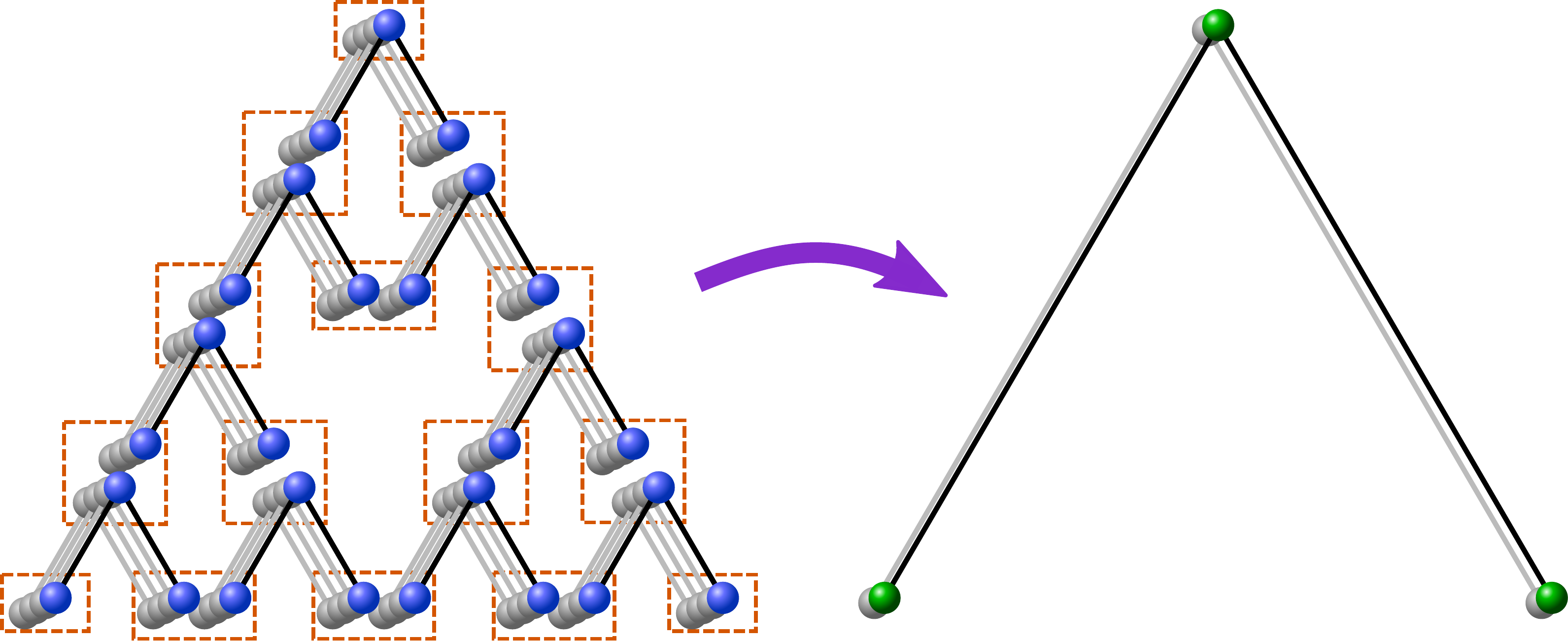}
 \caption{\label{fig:whole_setup} 2D setup for distributing a long-distance GHZ state on a triangular network using a scheme based on multpartite building blocks. The stations (signified by the orange boxes) implement the multipartite hashing protocol and merge the states in one step in a measurement-based way.}
\end{figure}

One way we can use the multipartite hashing repeater scheme is for distributing an entangled state over long distances. Exemplary we look at distributing a long-distance GHZ state with repeater stations arranged on a triangular grid as depicted in Fig. \ref{fig:whole_setup}, similar to the setup in \cite{2drepeater}.

\begin{figure}
 \includegraphics[width=\columnwidth]{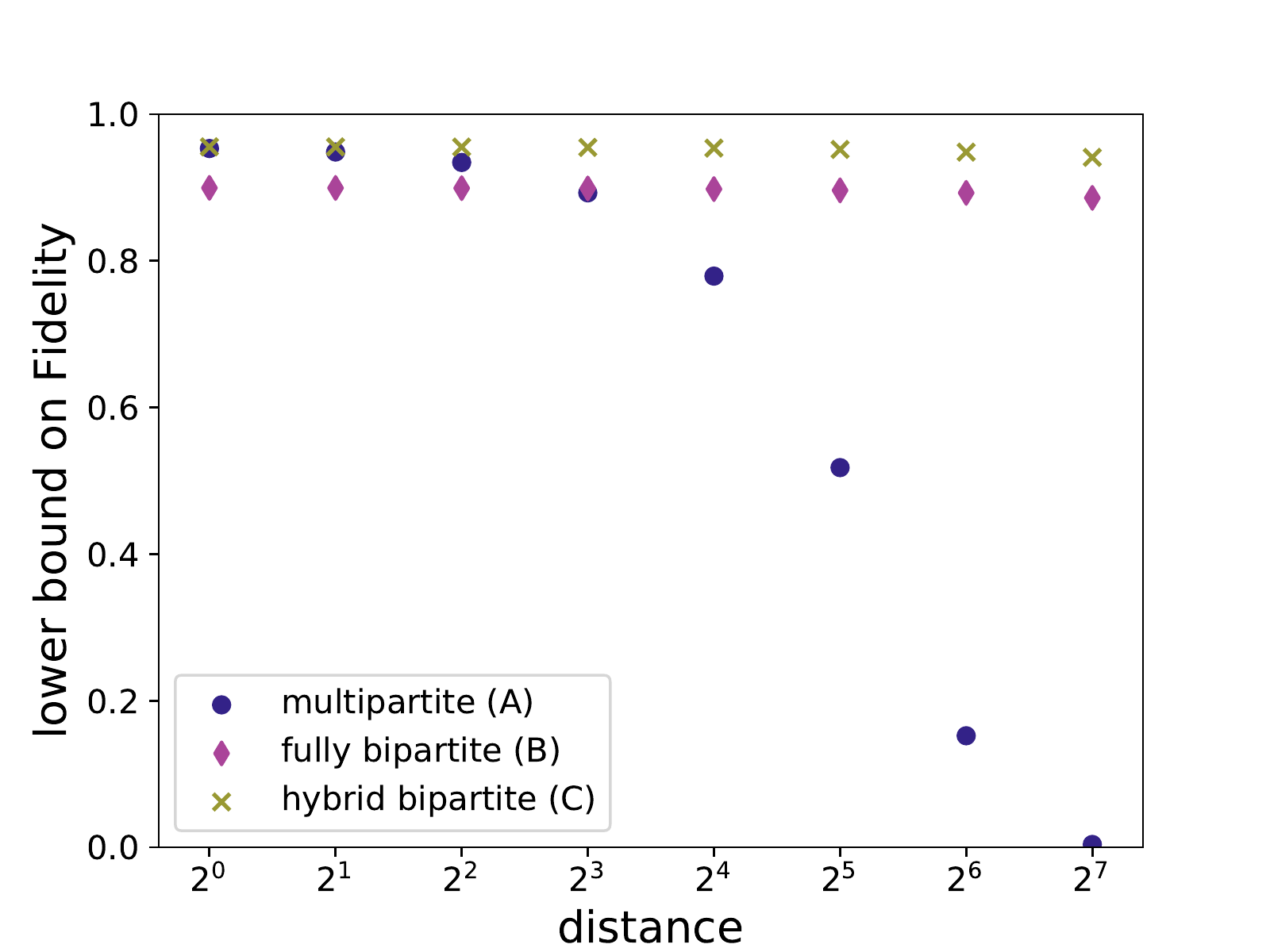}
 \caption{\label{fig:repeater_comparison} Comparing the reachable fidelities of $n\rightarrow1$ protocols for the multipartite scheme and the fully bipartite scheme for different numbers of repeater levels with a storage capacity of $1600$ qubits at each repeater station. The inital states are affected by local depolarizing noise with error parameter $q=0.99$ and the resource states with error parameter $p=0.98$ on each qubit.}
\end{figure}

While using the measurement-based implementation, which combines hashing and connecting the resulting states into a single step, asymptotically provides a scalable and deterministic protocol, in contrast to \cite{2drepeater} we cannot make use of the error detection process from using additional information gained at intermediate steps. Also, as we consider a scenario with limited storage and therefore the high amount of GHZ states used leads to a fidelity estimate by $F_\mathrm{GHZ}^{3^k}$ where $k = \log_2 L$ of the desired length $L$ in multiples of the elementary distance between stations on the triangular grid. In the bipartite (B) and hybrid (C) case one obtains $F_\mathrm{bip}^{2^{k+1}}$, which makes it clear that the advantage the multipartite approach (A) might have at small distances does not scale well for long distances.
In Fig. \ref{fig:repeater_comparison} one can see that the multipartite advantage is relevant for a short distances, but loses its advantage at longer distances.

\subsection{Cluster state \label{sec:cluster}}

Rather than distributing the same state over longer and longer distances, one can also consider a setup where the goal is to generate a state with a growing number of parties. To illustrate this mode of operation we consider building up large 2D and 3D cluster states from smaller building blocks.

\subsubsection{\label{sec:problem:blocks} Building blocks for cluster states}

In this section we present an approach for creating a 2D cluster state by merging smaller building blocks. We will arrive at two classes of building blocks, which we term windmill and shifter grid. They have in common that they reduce the required local storage capacity by a factor of two as compared to a strategy based on entangled pairs only. 

\begin{figure}
 \subfloat[\centering \label{fig:cluster_bell_pairs_gen}]{\includegraphics[width=0.48\columnwidth]{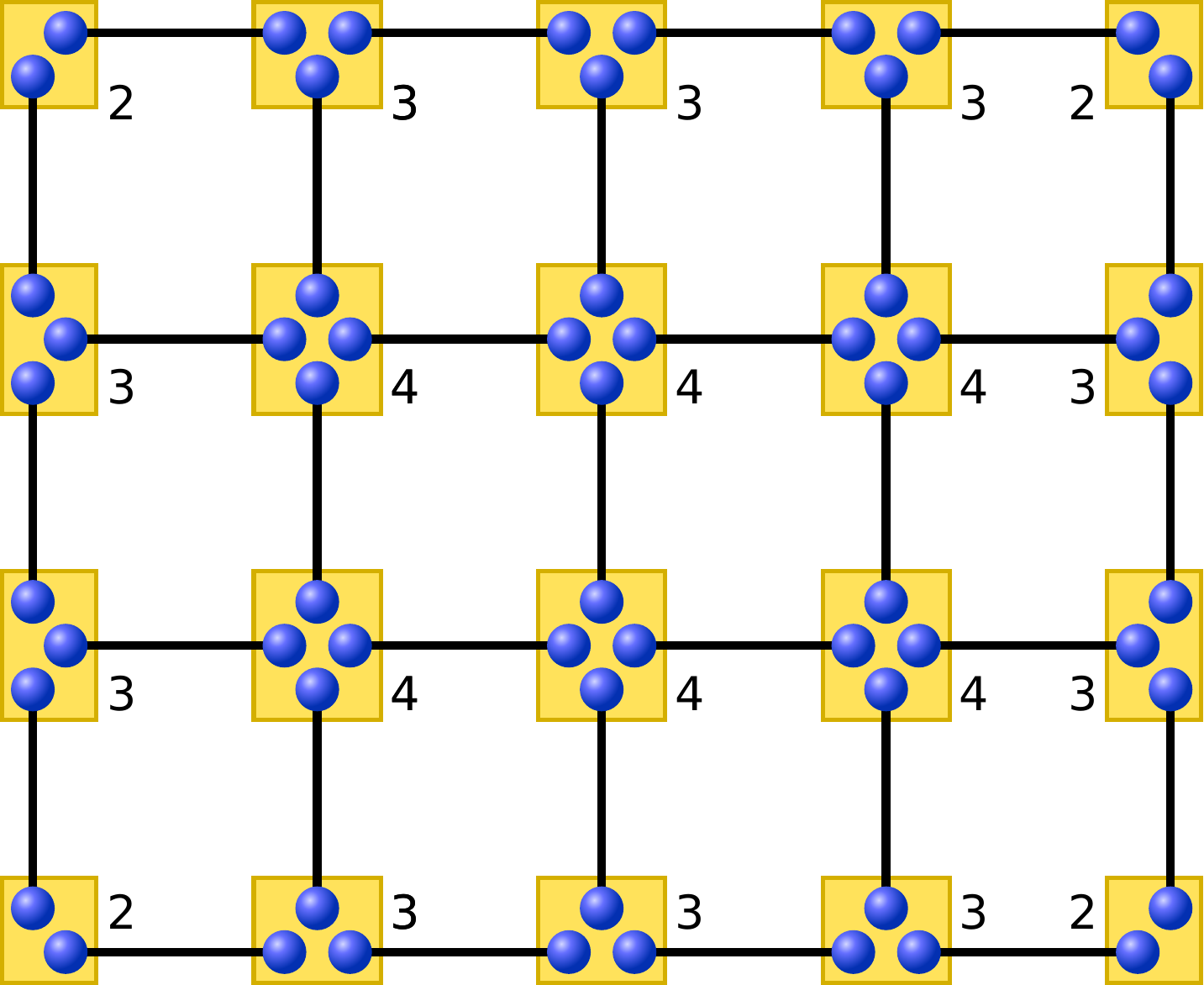}}
 \hfill
 \subfloat[\centering \label{fig:cluster_bell_pairs_gen_opt}]{\includegraphics[width=0.48\columnwidth]{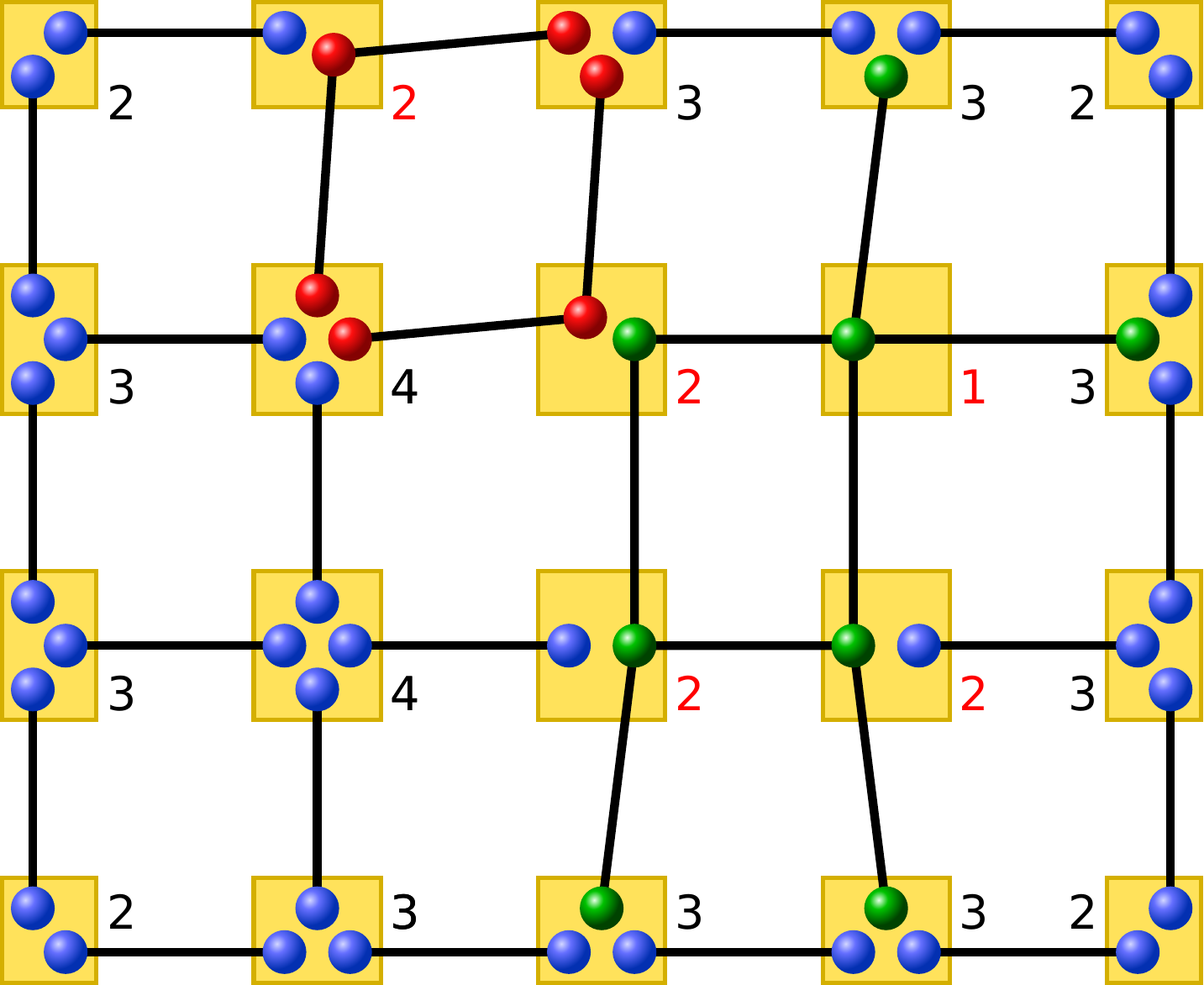}}
 \caption{(a) One can create a cluster state from a grid of Bell-pairs by merging them (orange rectangles). The numbers indicate how many qubits need to be stored at each of the stations. (b) Optimizing the storage required at the stations by using multipartite states: For the red qubits (GHZ-states) and the green qubits (2 colorable graph state) the entanglement distillation protocol for two colorable graph states are employed. The red numbers indicate where we have reduced the number of qubits which need to be stored at the station.}
\end{figure}

Let us start with a grid of Bell pairs as shown in Fig. \ref{fig:cluster_bell_pairs_gen}. If all stations merge their respective qubits we obtain a cluster state, which corresponds to the bipartite strategy. Note that at each node, four qubits need to be stored. That is, if a station is capable of storing $n$ qubits, the number of initial copies for the entanglement distillation protocol is $n/4$.

However, it is possible to consider different kinds of initial states to cover the whole grid such that combining the building blocks still results in a cluster state. In particular, any set of elementary building blocks that is obtained by merging some of the initial Bell-pairs in a grid can be used. Such covers of the cluster state, consisting of possibly several different substructures, can be highly irregular and one possible choice is shown in Fig. \ref{fig:cluster_bell_pairs_gen_opt}. Already from this example it is apparent that multipartite states can effectively reduce the storage capacity needed by the stations. \newline
Essentially, we have to identify covers that provide favorable storage requirements, ideally only needing to store two qubits at the borders of the elementary building blocks, so one can use $n/2$ initial copies for the multipartite entanglement purification protocol if each station can store $n$ qubits. This advantage is important as the fidelity of the output state after entanglement distillation strongly depends on the available number of initial states.\newline
We follow the idea of identifying possible configurations using two instead of four qubits at each station and we obtain two classes that form distinct blocks which are invariant under rotations with an angle of $\pi/2$. The idea is straightforward: one considers four neighboring stations, each storing four qubits belonging to Bell-pairs connecting them. Then, one merges these four qubits into two, thereby obtaining a small subgraph. By rotating this subgraph with an angle of $\pi/2$ four times one obtains a building block.

Figures \ref{fig:windmill} and \ref{fig:shifted_grid} show the \textit{windmill} and \textit{shifted grid} classes of building blocks as well as how these can be extended to larger building blocks, which have some central stations that only need to store one qubit per copy. These approaches can also be extended to 3D cluster states where we find coverings that use cubes instead of squares. The windmill blocks need to be modified to fit the 3D cluster but some stations will need to store $3$ qubits per copy as the neccessary dangling ends cannot be arranged in a better way. Here, the shifted grid approach shows that it scales very well to higher dimensions as the required graph states are simply cubes connected at each corner and only $2$ qubits per copy need to be stored at each station.

\subsubsection{Construction from smaller blocks \label{sec:fromblocks}}

\begin{figure}
\begin{minipage}{0.39\columnwidth}

 \subfloat[\centering]{\includegraphics[width=0.7\linewidth]{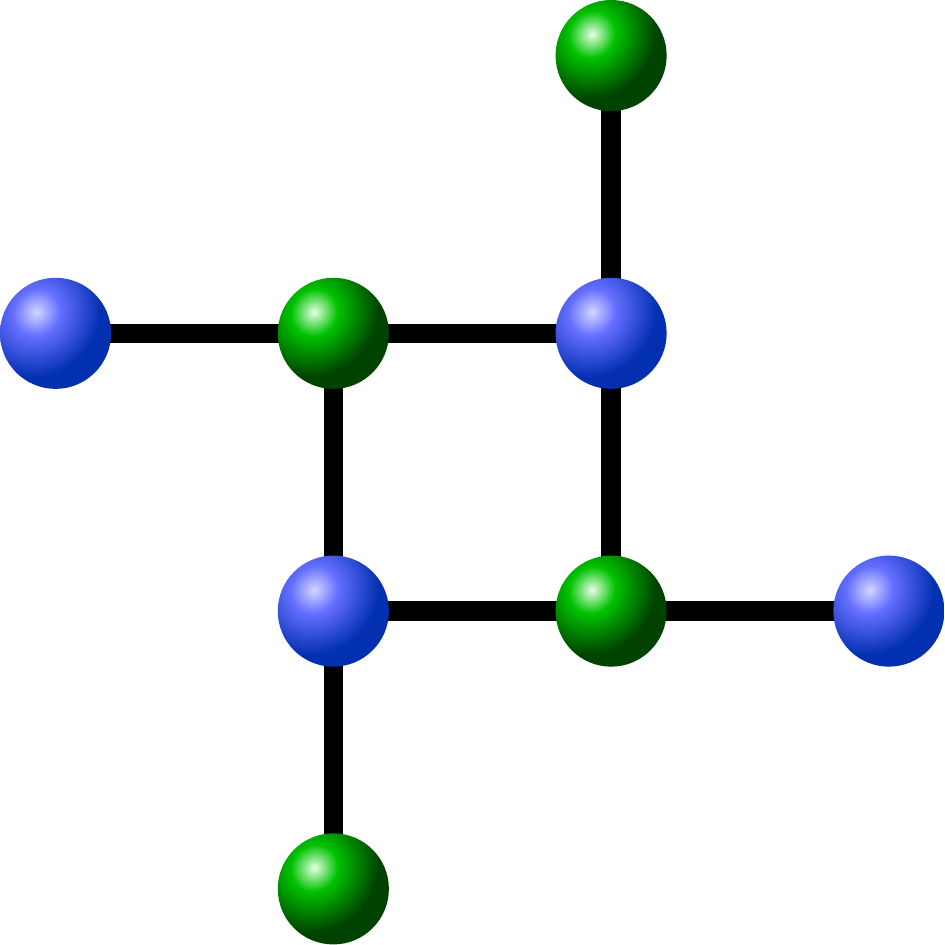}}

 \subfloat[\centering]{\includegraphics[width=0.8\linewidth]{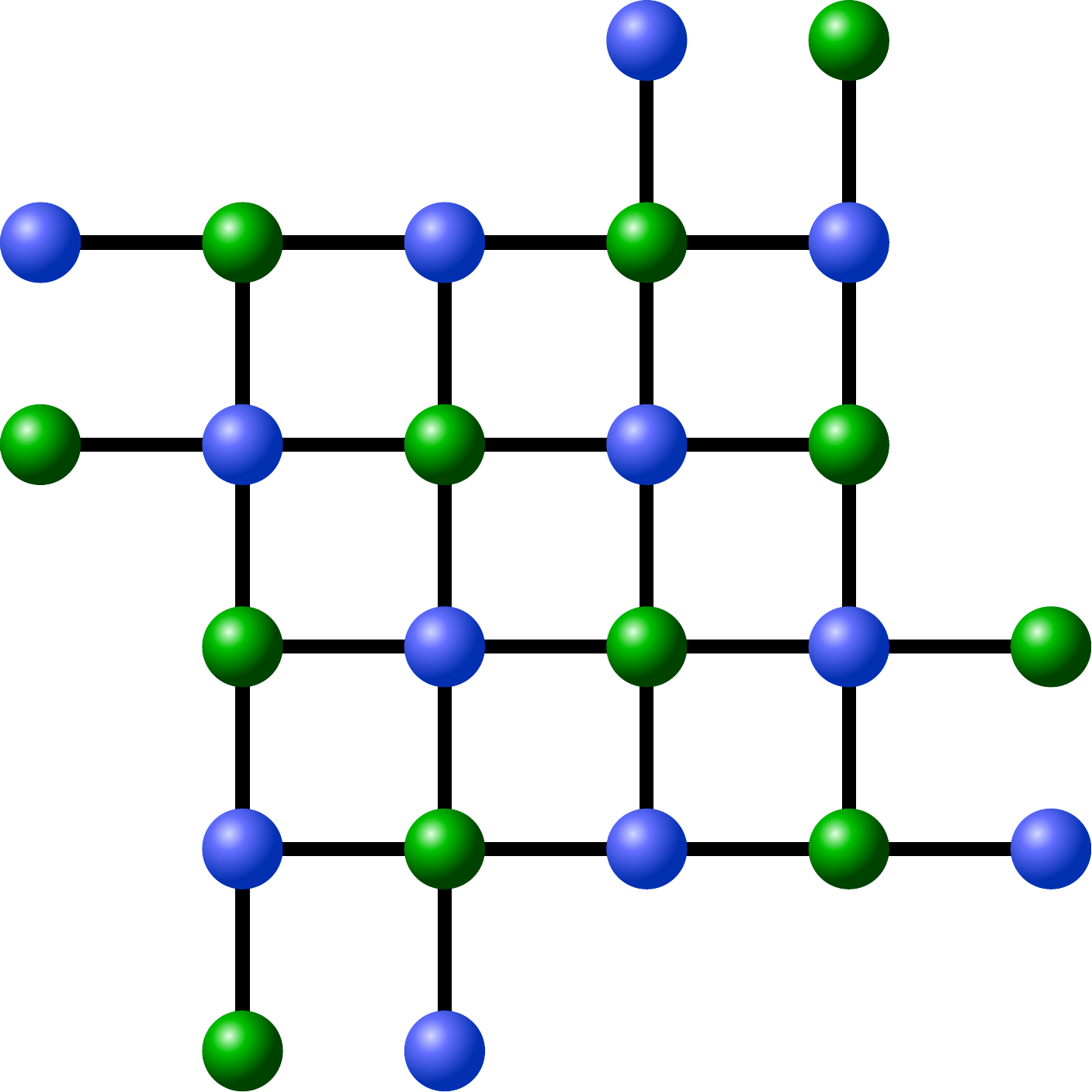}}

 \end{minipage} \hfill \begin{minipage}{0.59\columnwidth}
    \subfloat[\centering]{\includegraphics[width=0.9\linewidth]{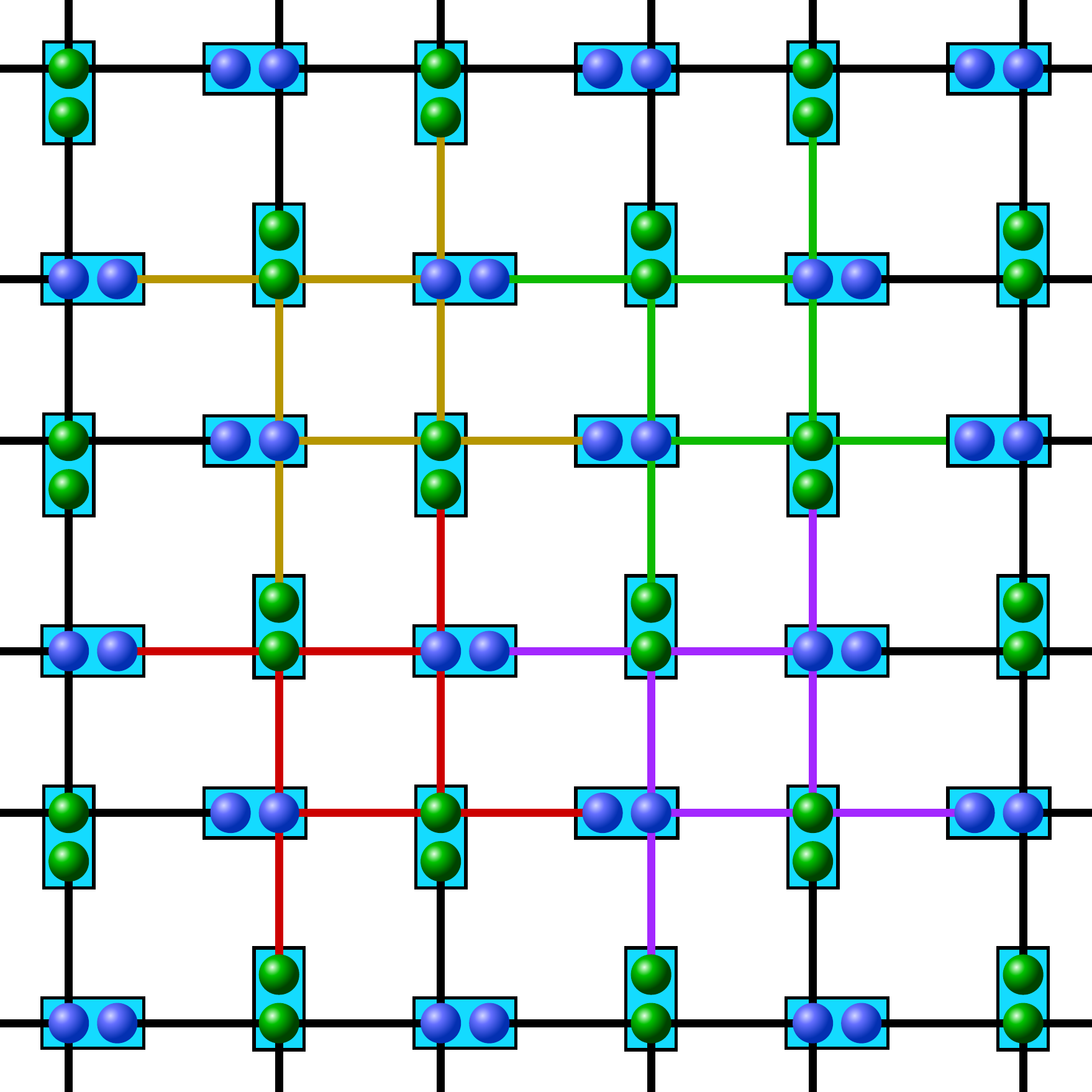}}
 \end{minipage}
 \caption{2D-cluster building blocks in a \textit{windmill} formation with a block size of (a) two or (b) four. (c) The blocks can be connected to a larger cluster state. The connection operations are performed on the qubits in the rectangles, which are stored at the same repeater station. Note that only two qubits per copy need to be stored at each location. \label{fig:windmill}}
\end{figure}

\begin{figure}
\begin{minipage}{0.39\columnwidth}
 \subfloat[\centering]{\includegraphics[width=0.5\linewidth]{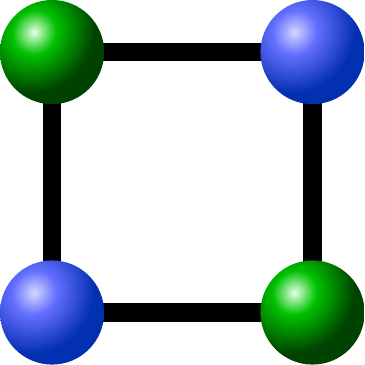}}

 \subfloat[\centering]{\includegraphics[width=0.8\linewidth]{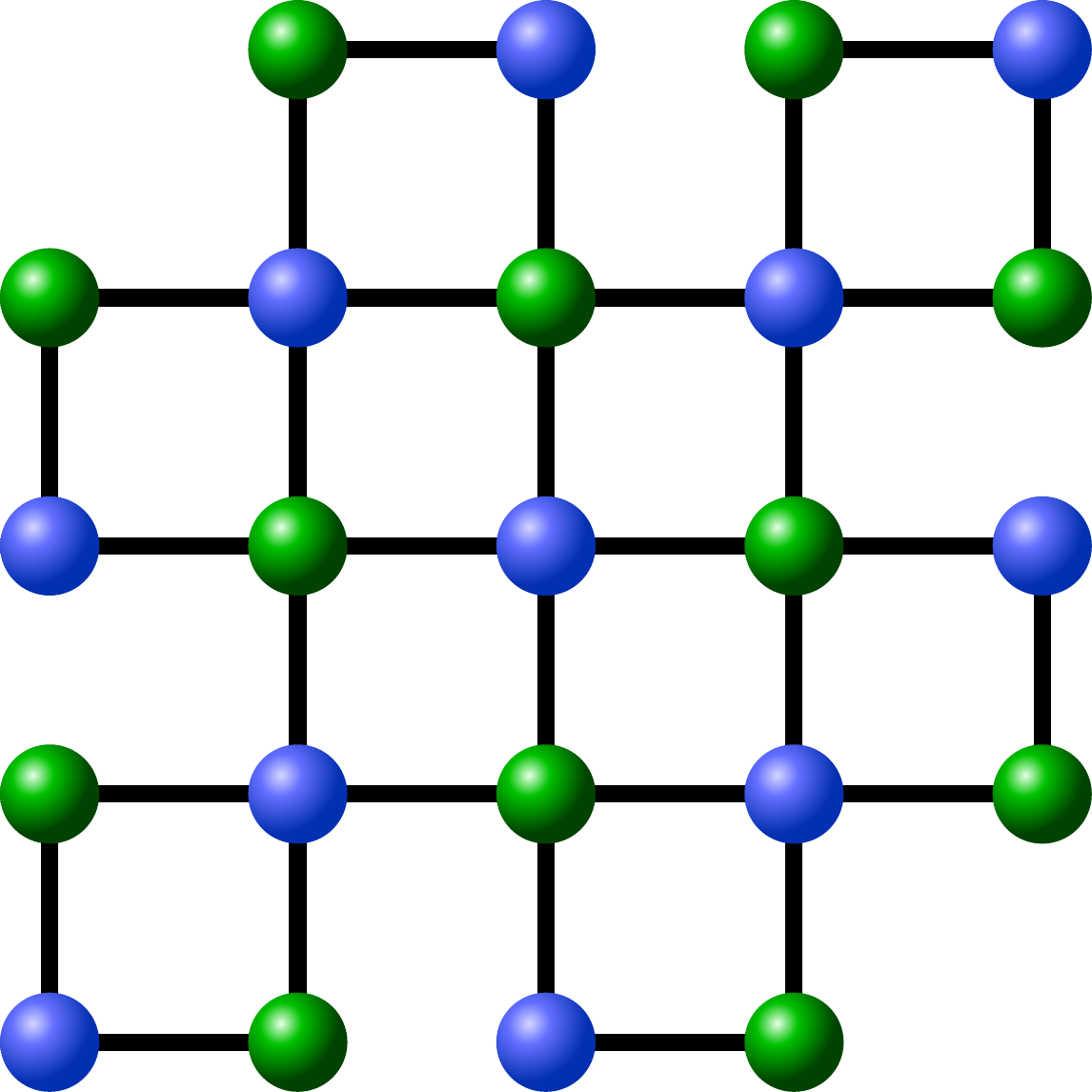}}
\end{minipage} \hfill \begin{minipage}{0.59\columnwidth}
 \subfloat[\centering]{\includegraphics[width=0.9\linewidth]{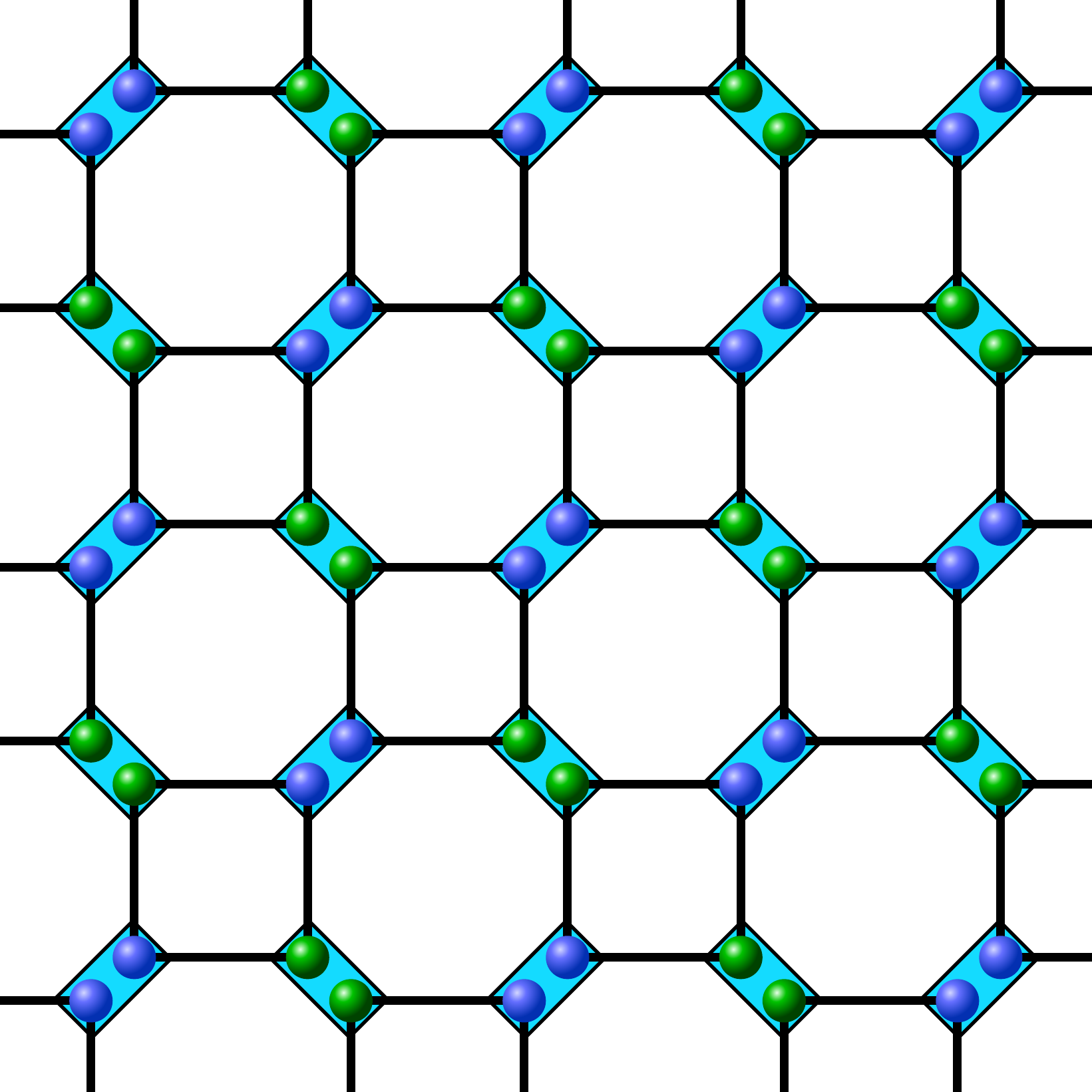}}
\end{minipage}
 \caption{2D-cluster building blocks in a \textit{shifted grid} formation with a block size of (a) one or (b) four. (c) The blocks are connected at the corners to form a larger cluster state. The connection operations are performed on the qubits in the rectangles, which are stored at the same location. \label{fig:shifted_grid}}
\end{figure}

We investigate using different schemes of constructing a 64x64 2D-cluster state with periodic boundary conditions from smaller building blocks. We compare the approaches using the windmill and shifted grid building blocks (Fig. \ref{fig:windmill} and \ref{fig:shifted_grid} respectively), which are both different variants of the multipartite approach (scheme A), and the bipartite approach (scheme B). Again, we use a straightforward error model of local depolarizing noise with error parameter $q$ acting on each qubit of the initial state. In this section we do not include noisy resource states as this only leads to a lower reachable fidelity for the purely bipartite scheme (B). Note that without imperfections in the resource states, schemes B and C are equivalent.

\begin{figure*}
 \centering
 \subfloat[\label{fig:2d_cluster_repeater} \centering]{\includegraphics[width=0.48\linewidth]{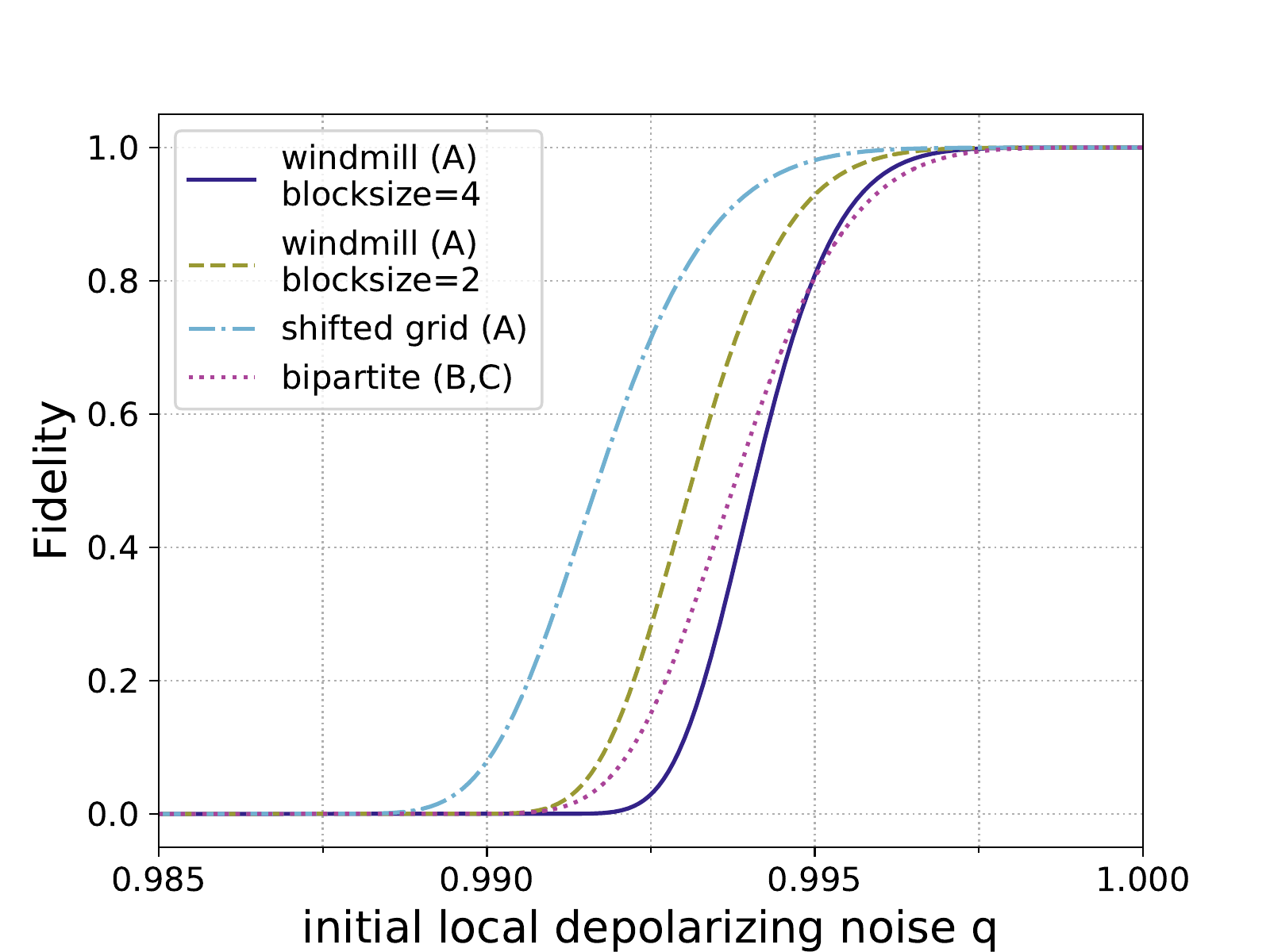}}
 \subfloat[\label{fig:2d_cluster_repeater_xy} \centering]{\includegraphics[width=0.48\linewidth]{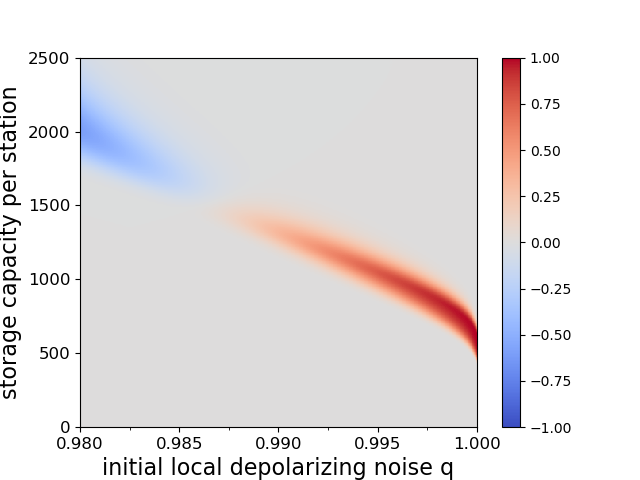}}

 \subfloat[\label{fig:2d_cluster_repeater_max_m} \centering]{\includegraphics[width=0.48\linewidth]{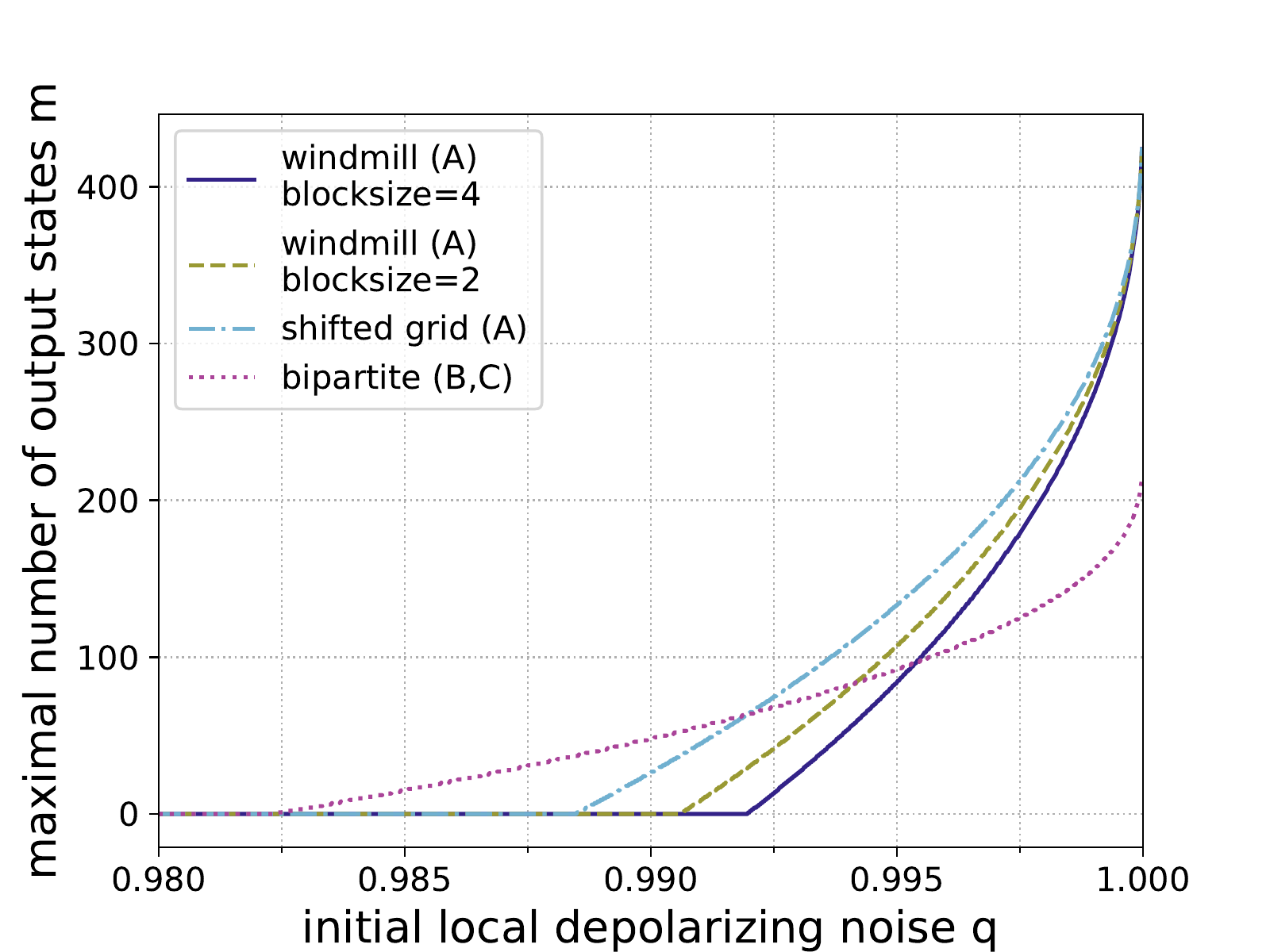}}
 \subfloat[\label{fig:2d_cluster_repeater_xy_max_m} \centering]{\includegraphics[width=0.48\linewidth]{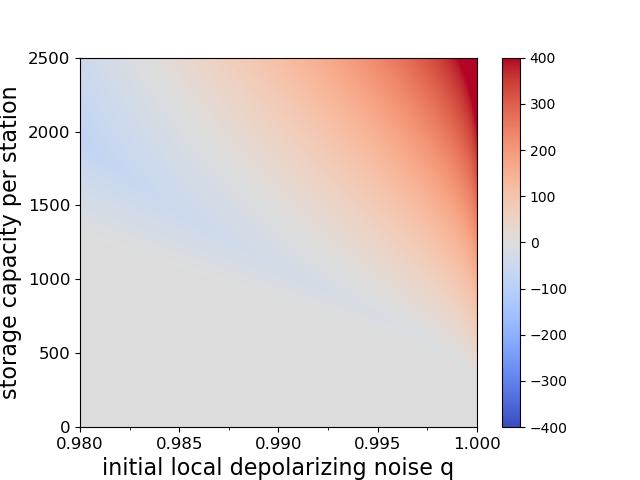}}
 \caption{Comparison of numerical results for generating $64 \times 64$ cluster states from smaller building blocks with hashing protocols using different architectures. (a) Reachable fidelities with local storage capacities of $1200$ qubits for $n\rightarrow100$ protocols. (b) Difference in fidelity $F_{\textrm{sg}} - F_\textrm{bip}$ between the shifted grid architecture $F_\textrm{sg}$ and the bipartite scheme $F_\textrm{bip}$ for $n\rightarrow100$ protocols. The red area signifies the parameter regime where the multipartite approach (A) achieves higher fidelities. The advantage of the multipartite approach for scenarios with low noise and very limited storage becomes apparent. (c) Number of output copies which can be provided with a fidelity of at least $0.9$ with local storage capacities of $1200$ qubits. Here the multipartite approach shows better scaling for low error rates. (d) Difference in obtainable output copies $m_{\textrm{sg}} - m_{\textrm{bip}}$ with a fidelity of at least $0.9$ by the shifted grid architecture $m_\textrm{sg}$ (A) and the bipartite approach $m_\textrm{bip}$ (B,C).}
\end{figure*}

First, let us consider a scenario where the storage capacity per location is limited. The achievable fidelities for a $n \rightarrow 100$ protocol are depicted in Fig. \ref{fig:2d_cluster_repeater} and \ref{fig:2d_cluster_repeater_xy}. While for the 2D cluster state the storage advantage is only a factor of two it is still very relevant.

Alternatively, instead of fixing the number of outputs, a practical question to ask is how many output pairs we can expect while still staying above a certain threshold fidelity. In figures \ref{fig:2d_cluster_repeater_max_m} and \ref{fig:2d_cluster_repeater_xy_max_m} the achievable number of output copies while keeping above a global threshold fidelity of $0.9$ is shown. Here it becomes clear that for $q$ close to $1$ the multipartite approaches can deliver more copies.

\begin{figure*}
 \centering
 \subfloat[\label{fig:3d_cluster_repeater} \centering]{\includegraphics[width=0.48\linewidth]{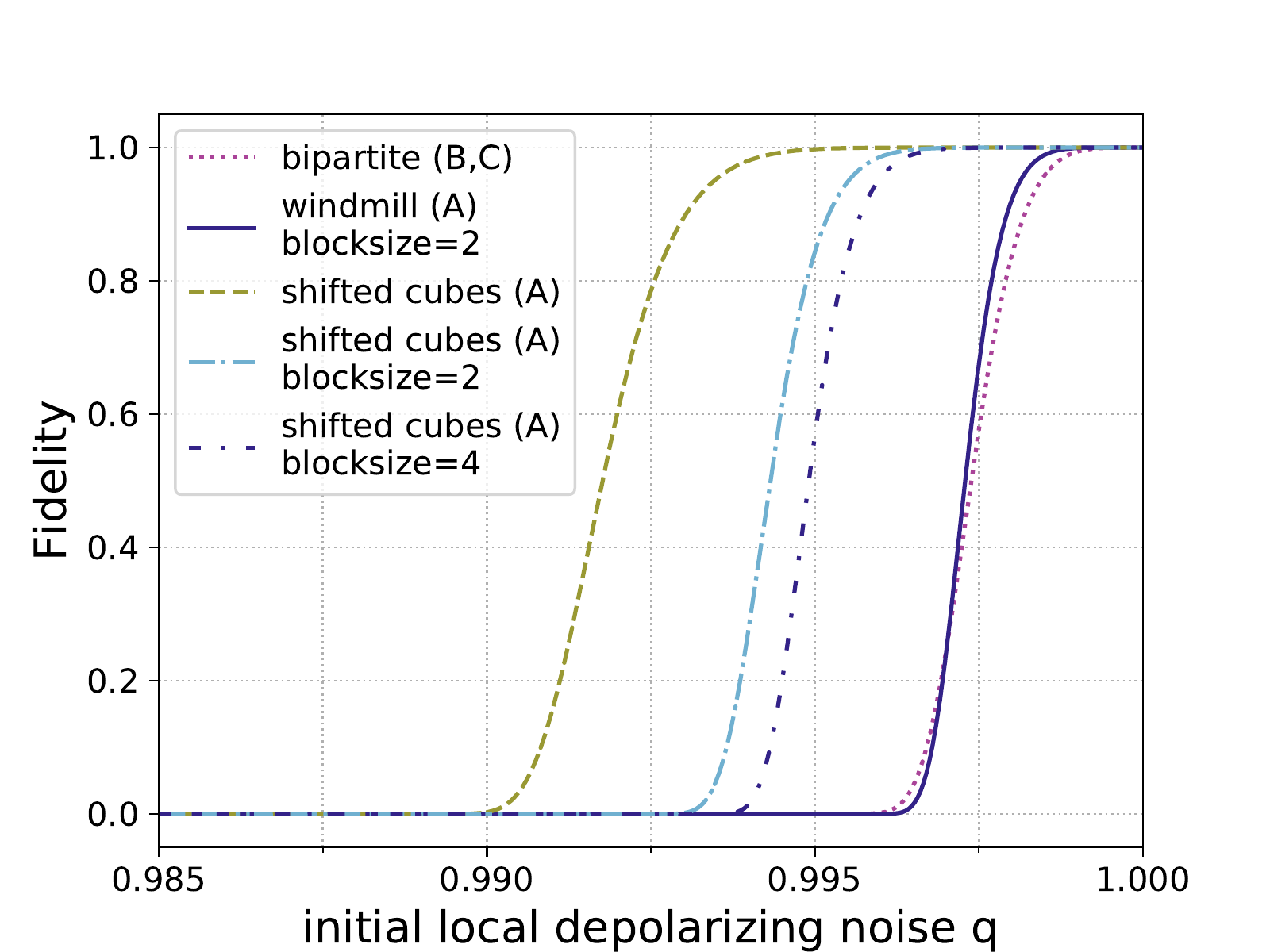}}
 \subfloat[\label{fig:3d_cluster_repeater_xy} \centering]{\includegraphics[width=0.48\linewidth]{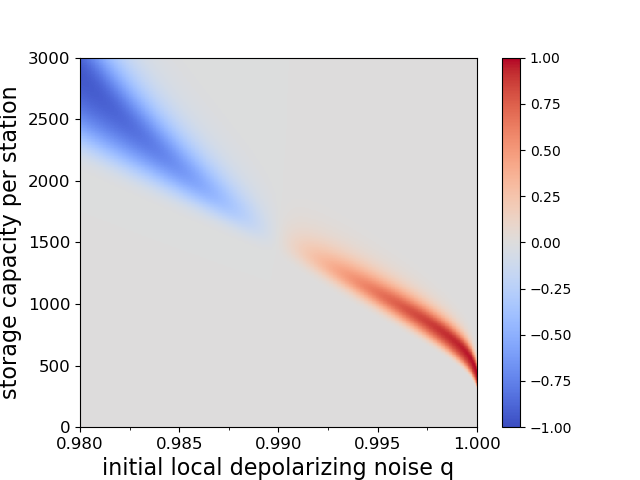}}

 \subfloat[\label{fig:3d_cluster_repeater_max_m} \centering]{\includegraphics[width=0.48\linewidth]{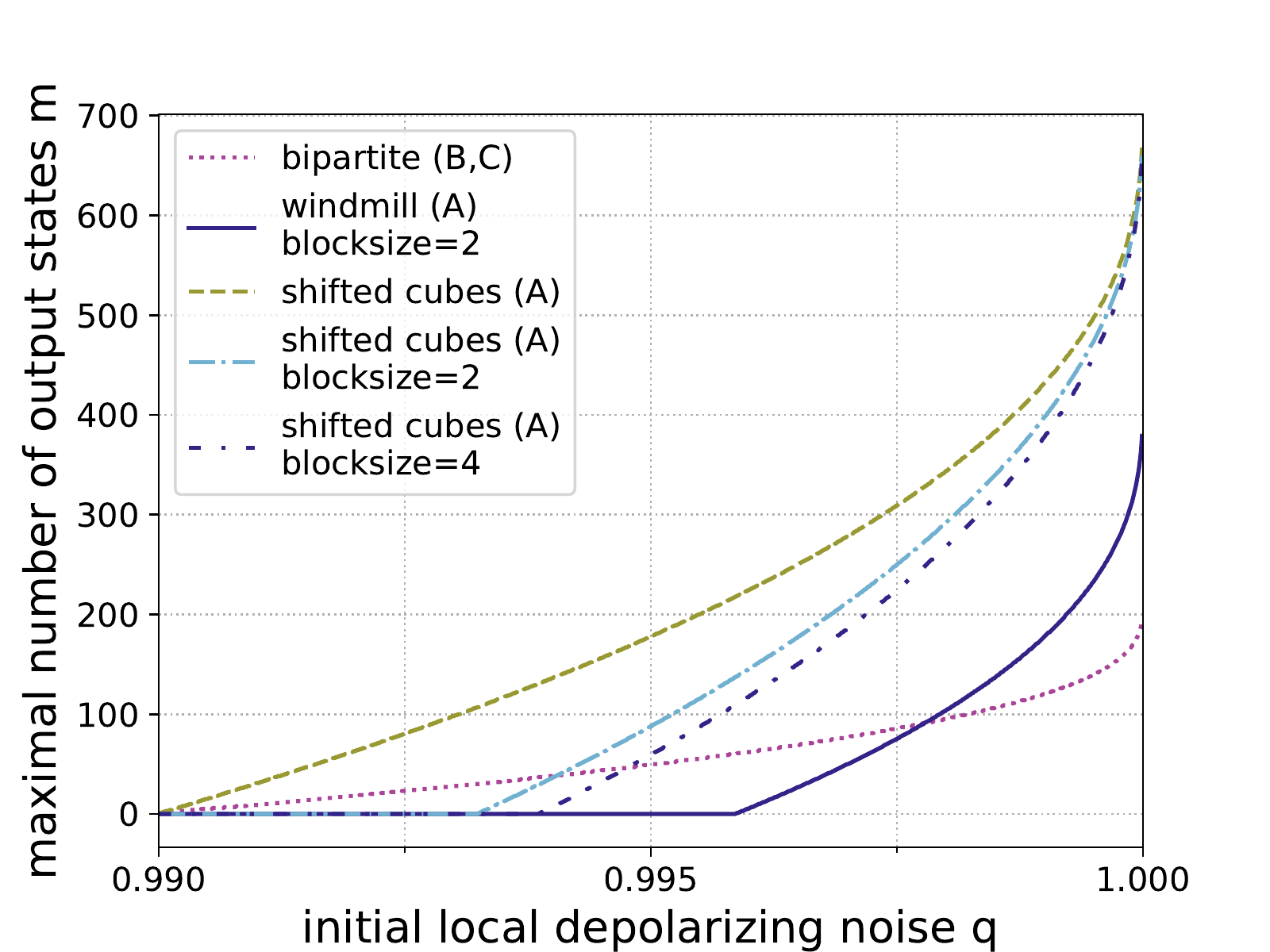}}
 \subfloat[\label{fig:3d_cluster_repeater_xy_max_m} \centering]{\includegraphics[width=0.48\linewidth]{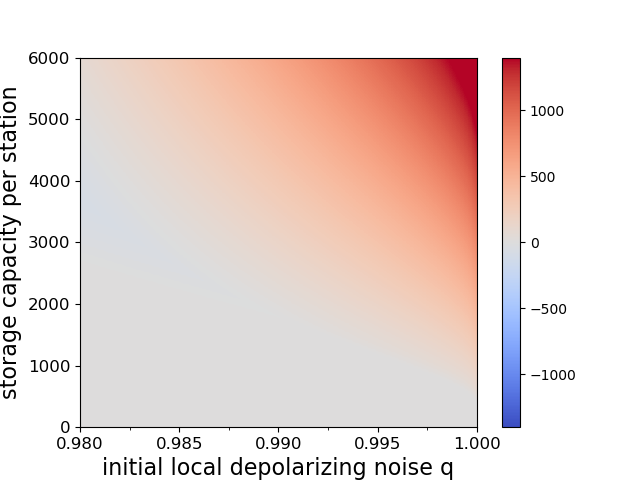}}
 \caption{\label{fig:all_3d_cluster_repeater}Comparison of numerical results for generating $64 \times 64 \times 64$ cluster states from smaller building blocks with hashing protocols using different architectures. (a) Reachable fidelities with local storage capacities of $1800$ qubits for $n\rightarrow100$ protocols. (b) Difference in fidelity $F_{\textrm{sg}} - F_\textrm{bip}$ between the shifted grid architecture $F_\textrm{sg}$ and the bipartite scheme $F_\textrm{bip}$ for $n\rightarrow100$ protocols. The red area signifies the parameter regime where the multipartite approach (A) achieves higher fidelities. (c) Number of output copies which can be provided with a fidelity of at least $0.9$ with local storage capacities of $1800$ qubits. Here the multipartite approach shows better scaling for low error rates. Difference in obtainable output copies $m_{\textrm{sg}} - m_{\textrm{bip}}$ with a fidelity of at least $0.9$ by the shifted grid architecture $m_\textrm{sg}$ (A) and the bipartite approach $m_\textrm{bip}$ (B,C).}
\end{figure*}

The same analysis is also extended to a 3D cluster states of size 64x64x64. See Fig. \ref{fig:all_3d_cluster_repeater} for the reachable fidelities and obtainable output copies in that case. Here the differences between the windmill and shifted grid blocks become more pronounced, as the shifted grid architecture allows to obtain a storage advantage of factor three while the windmill blocks only allow to store twice as many copies as in the bipartite approach. This is the reason why the shifted grid architecture performs very well in the three-dimensonal case.

\subsubsection{Building blocks with globally limited storage \label{sec:global_storage}}

\begin{figure}
 \centering
 \subfloat[\label{fig:2d_cluster_global_storage}\centering]{\includegraphics[width=0.8\linewidth]{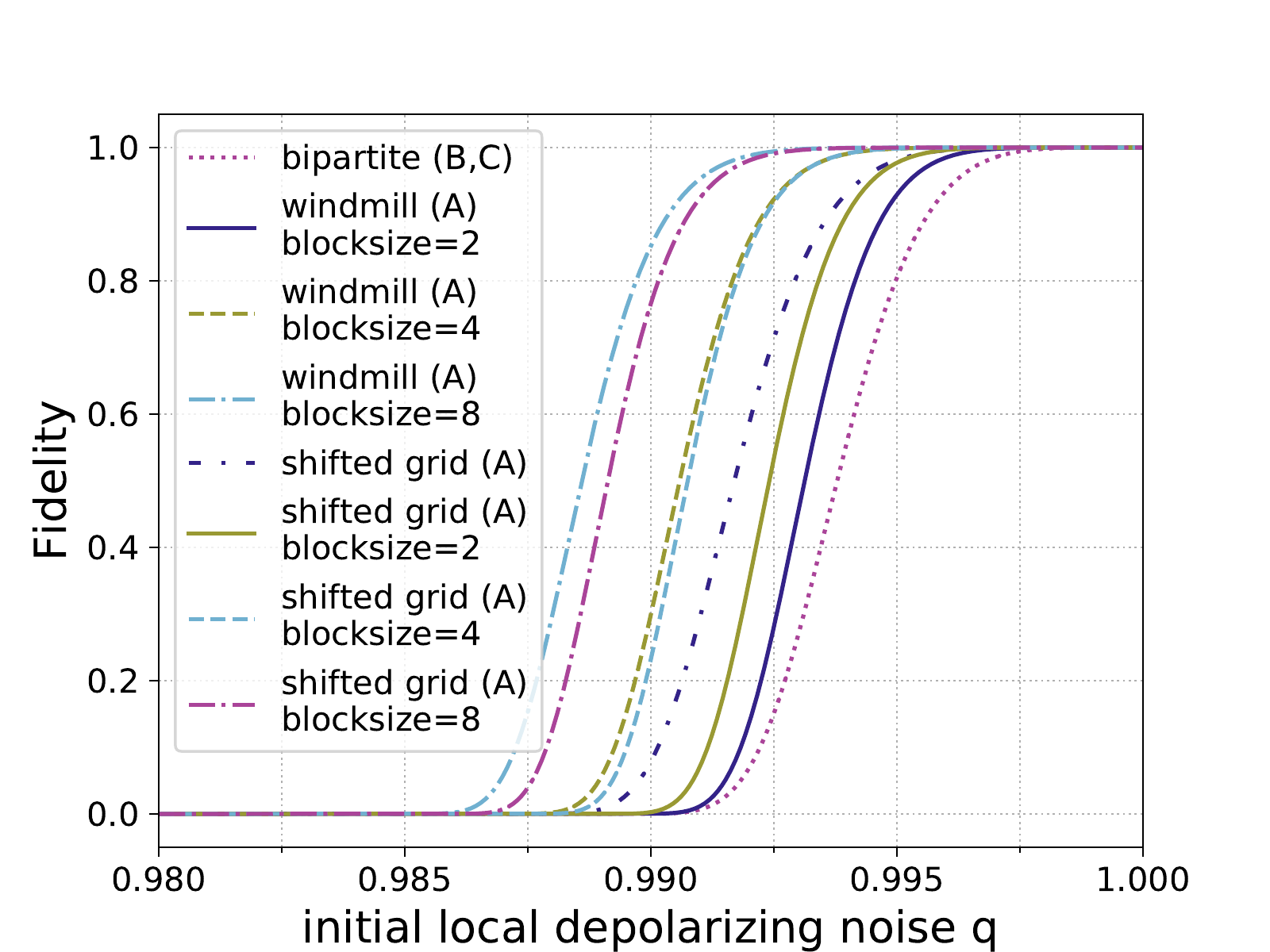}} \newline
 \subfloat[\label{fig:2d_cluster_global_storage_max_m}\centering]{\includegraphics[width=0.8\linewidth]{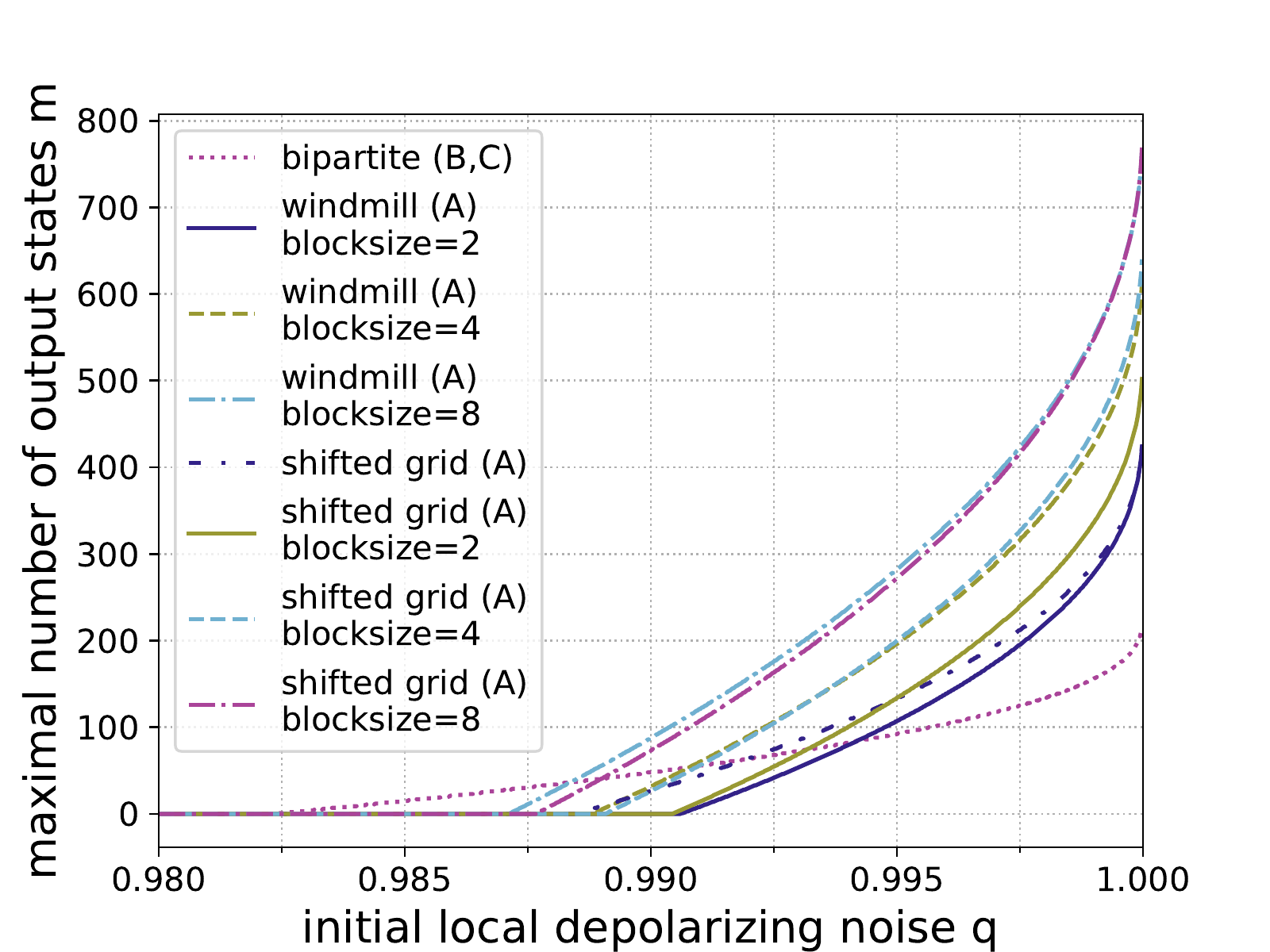}}
 \caption{\label{fig:all_2d_cluster_global_storage} Comparison of generating $64 \times 64$ cluster states from smaller building blocks when the storage capacity of $1200 \times 64^2$ qubits can be freely distributed among the involved stations. The multipartite approaches can profit from distributing the storage for more qubits to critical stations at the edges of the building blocks while the bipartite protocol cannot. (a) Reachable fidelities for $n\rightarrow100$ protocols. (b) Number of output copies that can be provided with a fidelity of at least $0.9$.}
\end{figure}

However, if one increases the block sizes, only the few repeater stations at the edges of the blocks need to store multiple qubits per copy. This means that if the system is limited by the total available storage rather than the storage per repeater station, the multipartite approach can benefit from this. This situation is akin to classical hard drives that are modular and can be moved to different servers depending on the requirements. Fig. \ref{fig:all_2d_cluster_global_storage} depicts the results for a two-dimensional cluster state. Here it becomes apparent that larger block sizes allow one to obtain an even bigger advantage. Another interpretation is that since the additional storage is only needed at the stations at the edge of the blocks, using a multipartite approach allows one to achieve the same or better result by only upgrading some of the locations with additional storage.

\begin{figure}
 \centering
 \subfloat[\label{fig:3d_cluster_global_storage}\centering]{\includegraphics[width=0.8\linewidth]{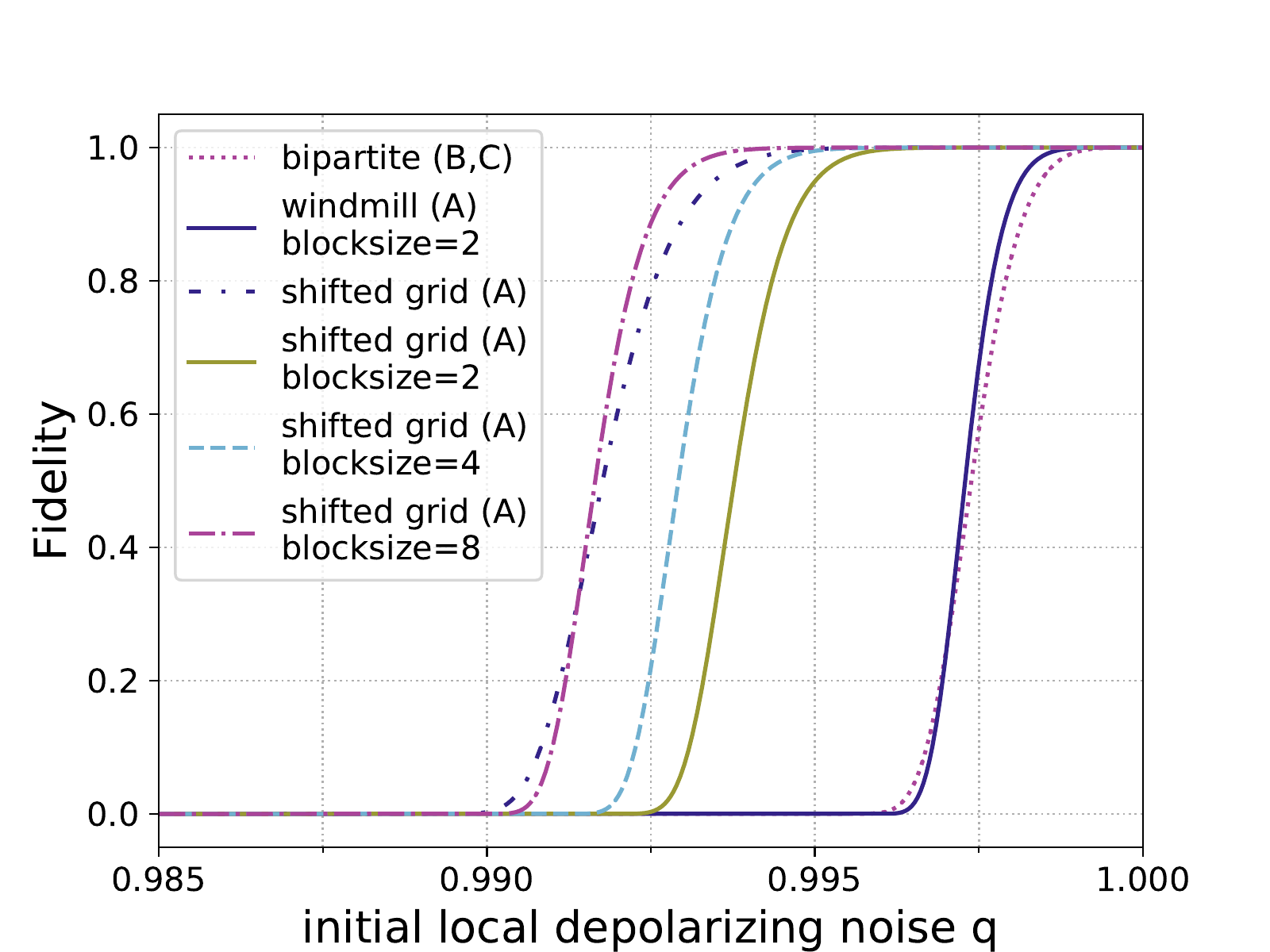}}\newline
 \subfloat[\label{fig:3d_cluster_global_storage_max_m}\centering]{\includegraphics[width=0.8\linewidth]{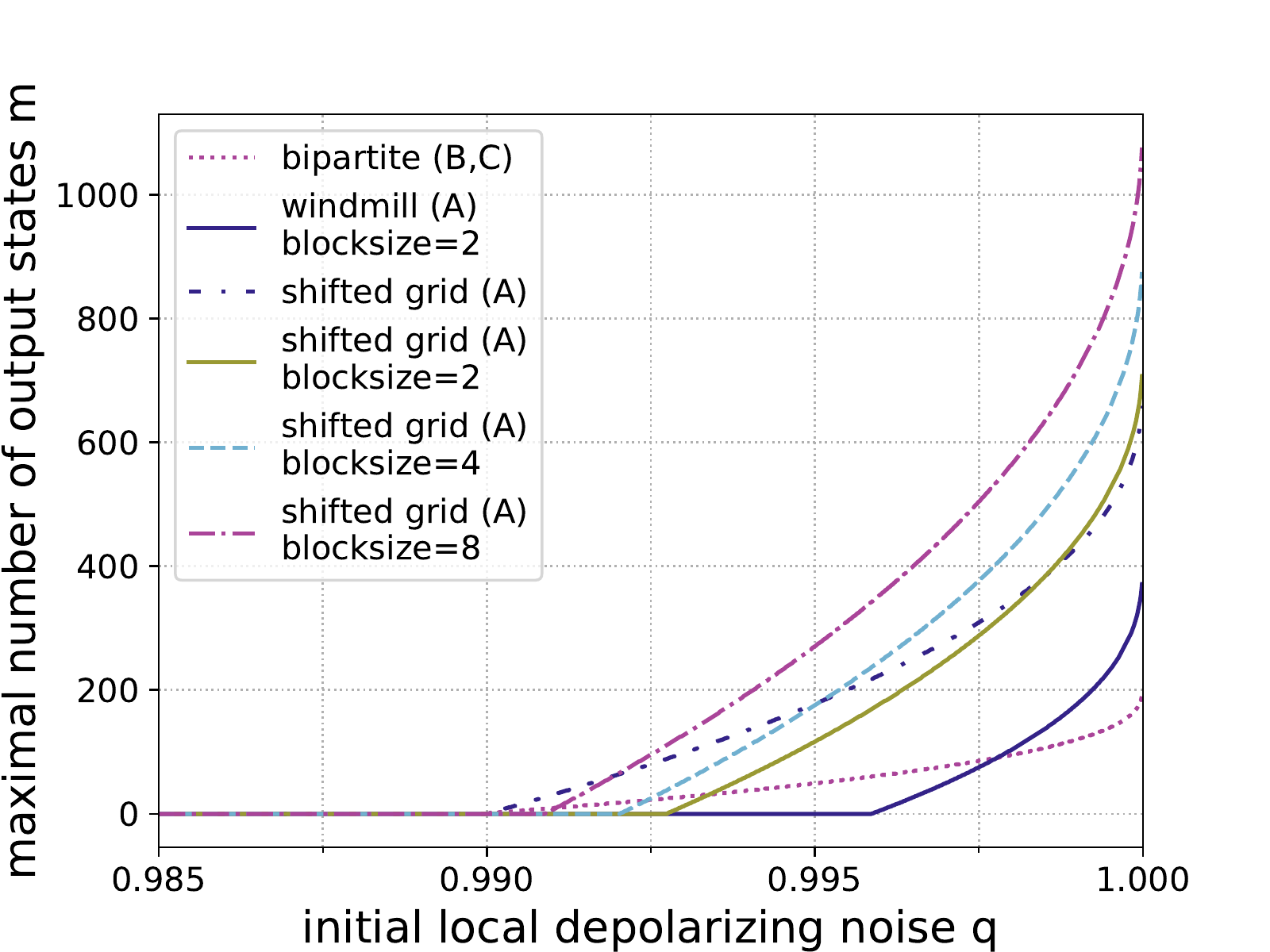}}
 \caption{\label{fig:all_3d_cluster_global_storage} Comparison of generating $64 \times 64 \times 64$ cluster states from smaller building blocks when the storage capacity of $1800 \times 64^3$ qubits can be freely distributed among the involved stations. (a) Reachable fidelities for $n\rightarrow100$ protocols. (b) Number of output copies that can be provided with a fidelity of at least $0.9$.}
\end{figure}

Similarly, Fig. \ref{fig:all_3d_cluster_global_storage} shows the results for the 3D cluster state when the storage capacity is limited globally. Interestingly, increasing the complexity of the building blocks actually makes them more vulnerable to noise at first as suddenly there are qubits with six neighbors in the input state. However, at larger block sizes the overwhelming storage advantage proves advantageous.

\subsubsection{Construction directly from Bell pairs \label{sec:frombell}}

Rather than relying on smaller building blocks, this model uses only Bell pairs with a noise model that has a clear physical interpretation. Bell pairs that are distributed between neighboring parties by sending one of the qubits through a noisy channel modeled by local depolarizing noise. We compare a bipartite approach where the Bell pairs are purified and then connected to a cluster state in the end and a multipartite approach where the Bell pairs are connected first, so only one qubit per site and copy needs to be stored, which consequently is purified using the multipartite hashing protocol. This intermediate step of storing the qubits is sensible for setups with imperfect storage capabilities because storing the resource state for the measurement-based implementation of the hashing protocol over long time periods is undesirable as any noise affecting the output qubits cannot be corrected.

The error pattern that arises from connecting these noisy Bell pairs where one qubit has been subject to local depolarizing noise with error parameter $q$ can be described by the noise channel:

\begin{equation}
 \begin{aligned}
  \mathcal{E}_{ab}(q) \rho = q \rho + \frac{1-q}{3} \Big( &Z^{(a)} \rho Z^{(a)} +Z^{(b)} \rho Z^{(b)} + \\
  + &Z^{(a)} Z^{(b)} \rho Z^{(b)} Z^{(a)} \Big)
 \end{aligned}
\end{equation}

acting on every edge of the cluster graph. See Appendix \ref{app:errorfrombell} for a detailed explanation. So the initial state for the multipartite entanglement distillation is given by:

\begin{equation}
 \prod_{\{a,b\} \in E} \mathcal{E}_{ab}(q) \Ketbra{G}{G}
 \label{eqn:frombell_final_noise}
\end{equation}
where $G$ is the graph associated with the cluster state.

\begin{figure*}
 \centering
 \subfloat[\centering \label{fig:cluster_from_bell}]{\includegraphics[width=0.32\linewidth]{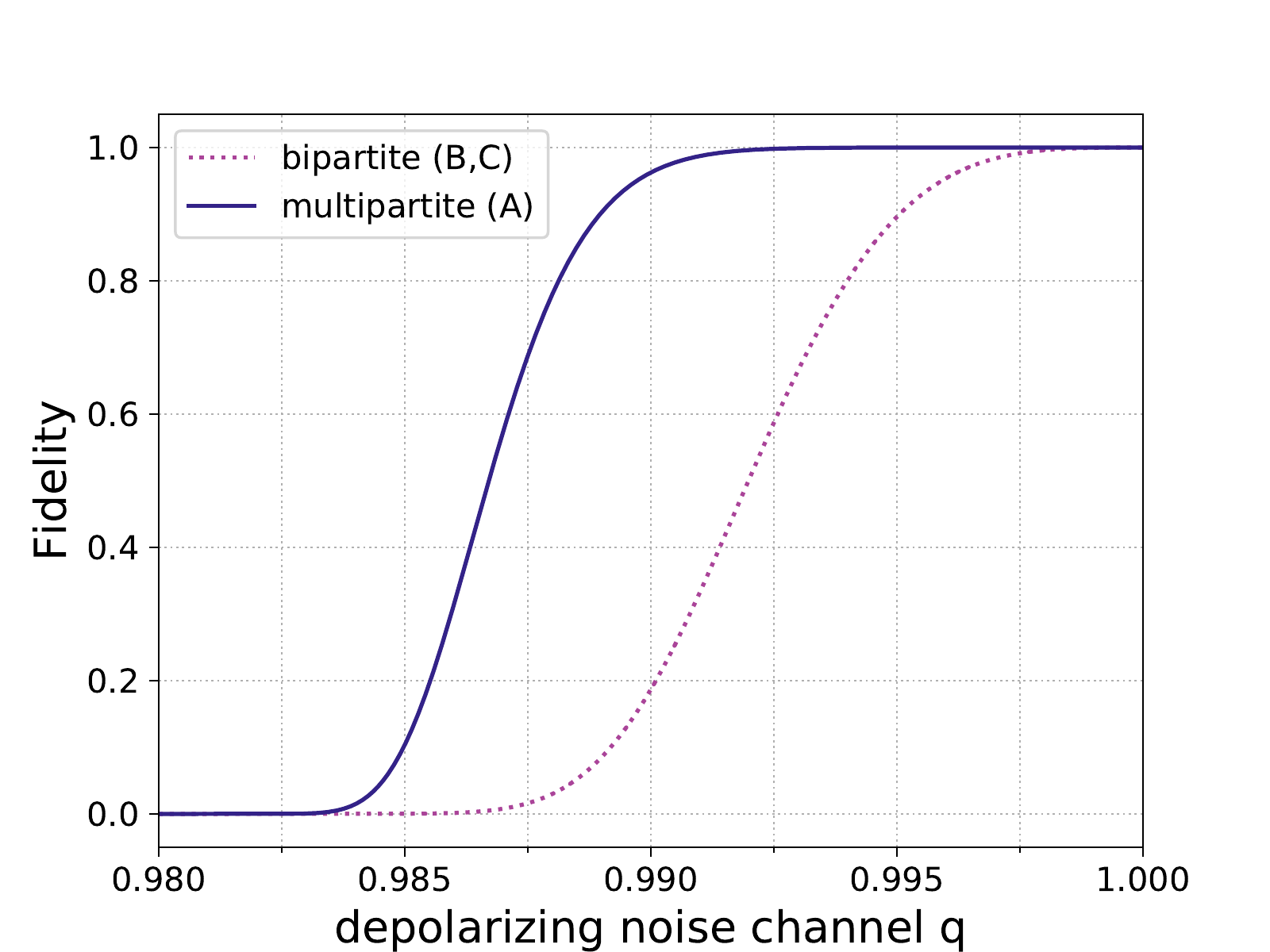}}
 \subfloat[\centering \label{fig:cluster_from_bell_xy}]{\includegraphics[width=0.32\linewidth]{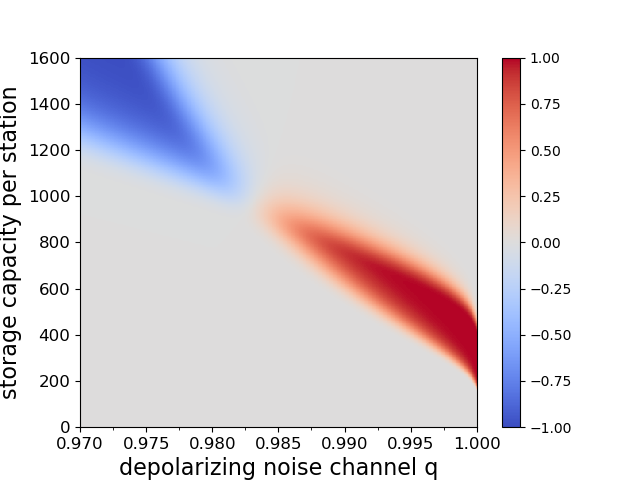}}
 \subfloat[\centering \label{fig:cluster_from_bell_max_m}]{\includegraphics[width=0.32\linewidth]{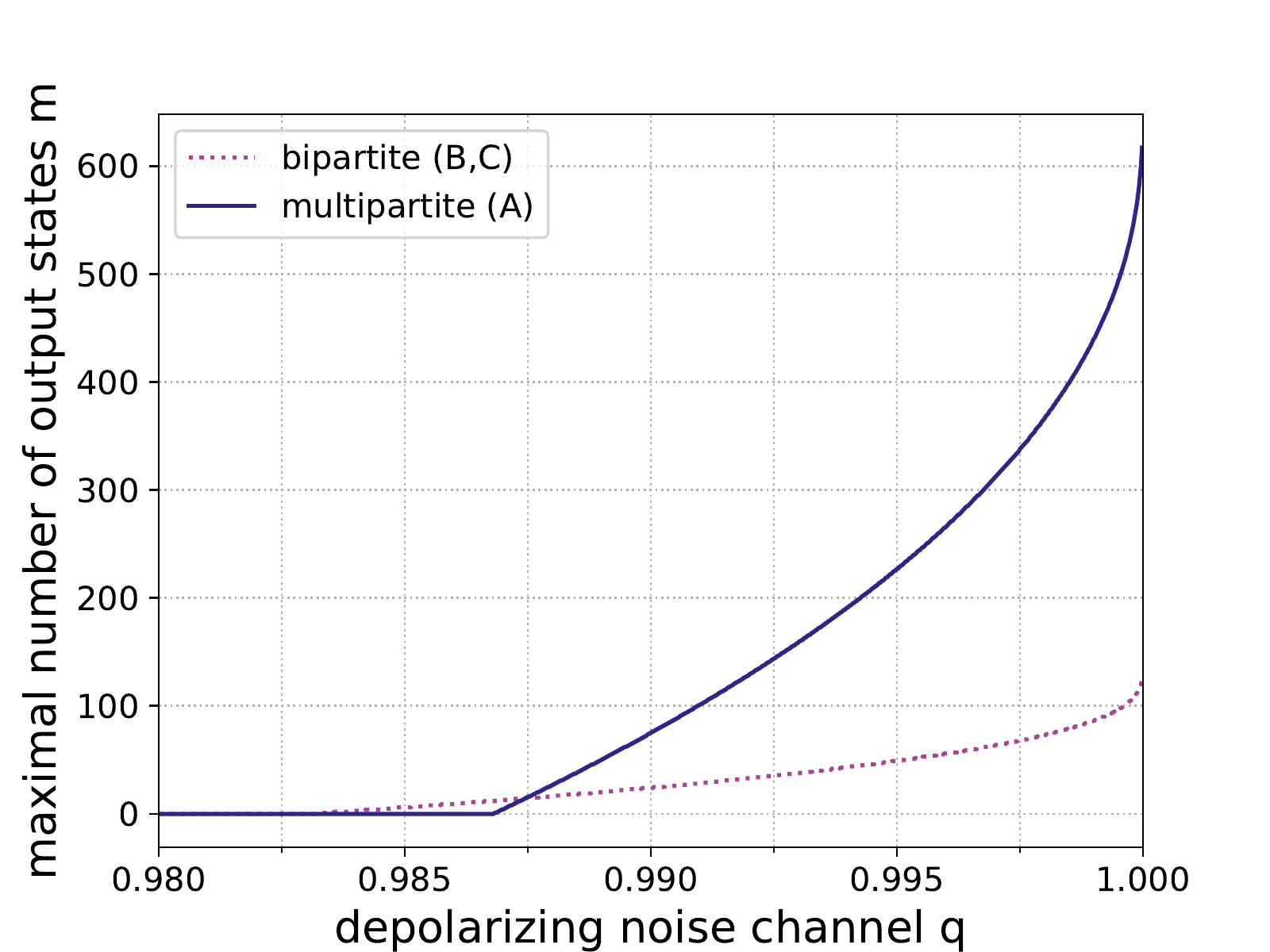}}
 \caption{\label{fig:all_cluster_from_bell} Comparison of multipartite (A) and bipartite (B,C) approach when distilling $64\times64$ cluster states generated directly from Bell pairs shared between adjacent parties where one half of the Bell pair is sent through a depolarizing noise channel with parameter $q$. (a) Reachable fidelity for $n\rightarrow50$ protocols with $800$ qubits storage capacity at each station. (b) Difference in fidelity between the multipartite and the bipartite architecture $f_\textrm{multi} - f_\textrm{bip}$. The red area signifies where the multipartite approach achieves higher fidelities. (c) Number of output copies which can be provided with a fidelity of at least $0.9$ with local storage capacities of $800$ qubits.}
\end{figure*}

In Fig. \ref{fig:all_cluster_from_bell} the results for both the reachable fidelity and obtainable output copies for a 2D cluster with dimensions $64\times64$ are depicted. One can clearly see that considering the multipartite approach for this scenario is also very relevant if storage capacities are limited.

\section{Arbitrary networks and generalization \label{Sec:generalization}}
The scheme for efficient generation of entangled states in a network that we have introduced and discussed for GHZ and cluster states can be generalized to other target states and network geometries. The key observation is that entanglement purification protocols for all graph states exist \cite{eppallgraphs}. This includes recurrence protocols, but also breeding and hashing protocols. The latter ones allow for entanglement purification of a large ensemble with constant yield, and can be implemented in a measurement-based way. The private fidelity \cite{zwerger_big_data} one can reach is arbitrarily close to one, approaching unit fidelity exponentially fast with the number of initial copies while the yield remains constant. This in turn allows one to connect and to merge arbitrary graph states in such a way that the final target state is generated with any desired target fidelity. It is essential that connection or merging processes are performed in the same step as the entanglement purification protocol, i.e. a single resource state implements both tasks. This leads to a drop in fidelity of the target state which can be lower bounded by $\prod F_i$, where $F_i$ are the (global) private fidelities of the connected or merged states, similarly as in the 1D case \cite{zwerger_big_data}. This can be compensated by using a logarithmically larger number of copies, where however the overhead per produced target state remains constant.

We start with the generalization of states with fixed number of qubits, but over larger distance as discussed in Sec. \ref{sec:GHZ} for three-particle GHZ states. In this case one merges and projects out short-distance states in such a way that a long-distance state of the same kind is produced, i.e. self-similar structures of growing scale are generated. This requires a regular network whose topology is associated with the desired target state.  This is called operational mode I in Ref. \cite{2drepeater}, where an example for the distribution of a 2D cluster-type state is described. The essential modification here is that hashing is used for entanglement purification for all states at once without nested levels, and is combined with the merging process in a measurement-based implementation. Some of the qubits are measured in this scheme, which is also combined with the purification process and implemented in a single step.

Also the scheme to generate cluster states shared between all nodes of a 2D square network can be generalized to networks with arbitrary geometry. This corresponds to operational mode II in Ref. \cite{2drepeater}, where again hashing is used for entanglement purification to obtain a scalable scheme with constant overhead. The goal is to generate a graph state that corresponds to the network structure. Consider as starting point a situation where Bell states are shared between all nodes of a network that are connected by edges. Notice that this is not the physical situation we consider, but rather used to illustrate the construction of elementary building blocks that are used. As discussed in Sec. \ref{sec:problem:blocks}, we can consider {\em any} merging of Bell pairs, regular or irregular, to generate elementary building blocks. These elementary building blocks are then purified and merged. All states generated in this way are graph states and can be purified via hashing. So for any choice of elementary building blocks, one obtains a scalable scheme with constant overhead, similarly as in the case of 2D and 3D cluster states. One can use small building blocks that are purified and merged, but also larger building blocks that are directly purified.

We finally remark that one can also consider the generation of graph states that do not correspond to the network geometry. There are two different graphs involved: a graph $G$ corresponding to the network geometry, and a graph $G'$ corresponding to the target graph state. There are multiple ways to generate target graph states. One strategy is to use all edges in the set $E \cap E'$, i.e. the direct links, and establish the missing edges $E' \backslash (E \cap E')$ by using a path formed by subsets of the edges of $E$. In particular, if we use the edges on such a path (which corresponds to a quantum channel) to establish short distance Bell-pairs, we can generate a virtual Bell-pair for that missing edge by performing entanglement swapping on all short distance Bell-pairs. These virtual Bell pairs can then merged again in an arbitrary way to form elementary building blocks \footnote{We remark that the elementary building blocks may also include vertices outside the vertex set $V'$ of the target graph state, e.g. to establish missing edges not in the set $E \cap E'$. For instance, if one wants to establish a 1D cluster state with additional edges of distance two in a 2D network, one can establish the additional edges by using nodes above and below the 1D line.}. Any choice of elementary building blocks with subsequent entanglement purification and merging leads to an efficient, scalable scheme with constant overhead per generated target state, independent of the size and distance of the states.

\section{\label{sec:conclusion} Conclusion and Outlook}
In this work, we introduced a repeater architecture for distributing multipartite entangled states in a quantum network with optimal scaling. The scheme is based on the multipartite quantum hashing protocol, where we make use of its fast convergence and favorable error thresholds in a measurement-based implementation. We have illustrated that the main elements that make quantum hashing an attractive tool for long distance point-to-point quantum communication \cite{zwerger_big_data}, namely constant overhead per node independent of the distance and non-zero yield, carry over directly to the application on multipartite graph states. Therefore, by applying this concept to all kinds of graph states we introduced a whole new class of protocols with optimal scaling. This includes the generation of long-distance states of few parties in regular networks, but also states shared between many or all parties by merging small elementary structures. A central element is to purify and merge in a single step using a measurement-based implementation, which leads to a scalable scheme with favourable error thresholds that enables one to transmit big quantum data.

We have also analyzed the performance in situations with limited resources. In particular, we considered situations where the global or local storage capacities are limited. In this case we found that using a multipartite or hybrid scheme offers advantages compared to approaches based on the distribution of entangled pairs using 1D repeaters. In particular, we constructed explicit schemes for generating 2D and 3D cluster states with minimal storage requirements per node, and also considered how to generalize this approach to a wide class of target states. The scheme we propose is based on the usage of a large number of copies, and hence requires a big infrastructure that might not be available in the near future.

We however believe that the analysis of network architectures with limited storage or other resources is of practical relevance also for near-term realizations of quantum networks. Also for networks with few nodes or small storage capacity, genuine multipartite approaches offer a storage advantage that might be harnessed. In this context, it would be particularly interesting to design network or repeater architectures for small-scale systems. This requires also the development of new, efficient entanglement purification protocols that operate on a small or medium number of copies, but offers similar advantages as the large-scale hashing protocols. We will report on such protocols elsewhere.

\section{Acknowledgements}
This work was supported by the Austrian Science Fund (FWF): P28000-N27 and P30937-N27

\bibliographystyle{apsrev4-1}
\bibliography{literature}

\newpage
\appendix

\section{Fidelity estimation for multipartite hashing with a finite number of input pairs \label{sec:multidetails}}
Continuing from the introduction in section \ref{sec:hashing}, we detail how the fidelity estimate for the multipartite quantum hashing protocol is obtained. This estimate is especially relevant in the case that we consider here, namely applying the protocol with a limited number of input copies. Bennet's inequality to estimate the fidelity of the bipartite hashing protocol was already used in \cite{pirker_secure_hashing, zwerger_big_data}. While error estimates of the multipartite protocol were present in those publications, the details of the fidelity estimate is the new aspect regarding the multipartite quantum hashing protocol presented in this paper.

The global fidelity of the hashing protocol can be estimated by looking at the probability that the initial states fullfil the requirements assumed in the formulation of the protocol. While this gets exponentially more likely as the number of input copies $n$ grows, we are interested in precisely the case where $n$ is relatively small. 

There are two error sources that might cause the hashing protocol to fail if $n$ is kept finite, which in turn decreases the fidelity. First, the bitstring $\widetilde{a}$ might fall outside the likely subspace. We use Bennett's inequality to bound this probability, i.e. the probability that the sample entropy $\widetilde{S}(\widetilde{a})$ differing from the von Neumann entropy $S(\rho)$ by more than $\delta$:
\begin{equation}
 \mathrm{P}\left( \left| S(\rho) - \widetilde{S}(\widetilde{a}) \right|  > \delta \right)   \leq    2 \exp \left\{ - n \frac{V}{a^2} h \left( \frac{a \delta}{V} \right) \right\}
\end{equation}
with $a=\max_i \left| - \log_2 \lambda_i - S(\rho) \right|$, $V = \sum_i \lambda_i \log_2^2 \lambda_i - S^2(\rho)$ and $h(u) = (1+u) \log (1+u)$. For sufficiently high n this expression scales like $\alpha \exp{(-\beta n \delta^2)}$ for some $\alpha$ and $\beta$ that depend on the input states.

Second, it might not be possible to uniquely identify $\widetilde{a}$ after a finite number of parity measurements, even if it falls into the likely subspace. The probability that any two different bitstrings show the same result after measuring a random subset parity is $1/2$. From the noiseless coding theorem we know that the likely subspace contains at most $2^{n(S(\rho) + \delta)}$ states, the probability of not uniquely identifying $\widetilde{a}$ after $k$ steps (consuming $k$ input states) is smaller than $2^{n(S(\rho) + \delta) - k}$.

We choose $k = n (S(\rho) + 2 \delta)$, as suggested in \cite{bennett_hashing}, and obtain for the success probability, i.e. the global fidelity of all output states being in the desired state:

\begin{equation}
 F \geq 1 - 2 \exp \left\{ - n \frac{V}{a^2} h \left( \frac{a \delta}{V} \right) \right\} - 2^{-n\delta} = f(\widetilde{a}, n, \delta)
\end{equation}

Note that in the scenarios we consider one chooses the number of input qubits $n$ and output qubits $m$ and derives $\delta$:
\begin{equation}
 \delta = 1/2 \left( 1 - S(\rho) - m/n \right)
\end{equation}
$\delta > 0$ has to be met in order for the protocol to be viable with the given parameters, which also directly imposes a condition on possible values of $m$.

Now, as mentioned in the main part, for multipartite states a separate subprotocol is needed for each color of the graph and one bitstring per qubit has to be considered separately.
The relevant entropy for the $k$-th qubit substring is given by $S_k$:
\begin{equation}
 S_k = S(a^{(k)}) = - \lambda_{k,0} \log_2 \lambda_{k,0} - \lambda_{k,1} \log_2 \lambda_{k,1}
\end{equation}
where $\lambda_{k,i} = \sum_{\mu_{j \neq k}} \lambda_{\mu_1 \dots \mu_{k-1} i \mu_{k+1} \dots \mu_N}$ is the probability that the $k$-th bit in the graph state basis vector $\bm{\mu}$ equals $i$. 

For two-colorable graph states we define $S_A = \max_{k \in A} S_k$ and $S_B = \max_{k \in B} S_k$ for the vertices with colors $A$ or $B$, respectively. Because $m/n = 1 - S_A - S_B - 2 \delta_A - 2 \delta_B$, only the sum of $\delta_A$ and $\delta_B$ is fixed, and we have to choose how to distribute these additional pairs between the subprotocols which can have an important effect for some error types (see section \ref{sec:restricted_error} ). The asymptotic yield is given by $m/n = 1 - S_A - S_B$ as $\delta_A$ and $\delta_B$ are allowed to approach $0$ as $n$ tends to infinity.

We define 
\begin{equation}
f_k = 1 - 2 \exp \left\{ - n \frac{V_k}{a_k^2} h \left( \frac{a_k \delta_k}{V_k} \right) \right\} - 2^{-n\delta_k}
\end{equation}

where $a_k = \max_i \left| - \log_2 \lambda_{k,i} - S_k \right|$ and $V_k = - \lambda_{k,0} \log_2 \lambda_{k,0} - \lambda_{k,1} \log_2 \lambda_{k,1}$. If the noise is symmetric such that all $S_k$ are equal for $k \in A$ then also $\delta_k = \delta_A$ have the same value. However, since all the bitstrings for one color can be evaluated simultaneously, a smaller $S_k$ means that for that $k$ $\delta_k$ is automatically chosen larger as even more copies are used. It holds that $\delta_k = \delta_A + (S_A - S_k)/2$ for $k \in A$. An anologous rule is derived for color $B$.

Finally, the global fidelity can be estimated by simply multiplying the success probabilites for the individual strings:

\begin{equation}
 F \geq \prod_{k \in V} f_k
\end{equation}

\section{Error model from connected Bell pairs \label{app:errorfrombell}}
In Sec. \ref{sec:frombell} we discuss a model based on Bell pairs where one qubit has been affected by local depolarizing noise with error parameter $q$ (as defined in \eqref{eqn:ldn_def}). These Bell pairs are then connected to a two-dimensional cluster state. In this section we derive the error pattern on that cluster state which arises from the noise on the Bell pairs.

One central property of the graph state basis is that any Pauli-diagonal noise channel can be written as a combination of correlated $Z$-noises (see e.g. \cite{He06}). We consider the graph state corresponding to the graph of two connected vertices $1$ and $2$ that can be written as $\ket{G_\mathrm{bip}}^{(1,2)} = 1/\sqrt{2} \left( \ket{0}_1 \ket{+}_2 + \ket{1}_1 \ket{-}_2 \right)$, which clearly is local-Clifford equivalent to the standard Bell states. Local depolarizing noise acting on one qubit of that graph state can be described as: 

\begin{equation}
 q \mu + \frac{1-q}{3} \left( Z^{(1)} \mu Z^{(1)} +Z^{(2)} \mu Z^{(2)} + Z^{(1)} Z^{(2)} \mu Z^{(2)} Z^{(1)} \right)
\end{equation}
where $\mu = \ketbra{G_\mathrm{bip}}{G_\mathrm{bip}}^{(1,2)}$.

Now we consider the situation where that state is connected to some other graph state $\Ket{G^\prime}$ with the connection procedure described in Sec. \ref{sec:graphconnect} using the qubit that we label $3$ as the point where the bipartite graph state will be attached to $\Ket{G^\prime}$.
The connection operation is given by $\mathrm{CNOT}^{2 \rightarrow 3} = \ketbra{0}{0}^{(2)} \otimes \mathbbm{1}^{(3)} + \ketbra{1}{1}^{(2)} \otimes X^{(3)}$ followed by a $Z$-measurement on qubit $3$. Applying $\mathrm{CNOT}^{2 \rightarrow 3}$ transforms the separate graph states to a new graph state corresponding to a graph where the new neighborhood $N^\prime_2$ of qubit $2$ is given by $N_2 \bigcup N_3 - N_2 \bigcap N_3$.

First, we investigate what effect the noise on $\ket{G_\mathrm{bip}}$ has on the resulting state. $Z^{(2)}$ commutes with $\mathrm{CNOT}^{2 \rightarrow 3}$ and also has no bearing on the outcome of the $Z$-measurement on qubit $3$. Furthermore, with the possible local-Clifford corrections depending on the measurement outcome consist only of applications of $Z$ to various qubits \cite{He06}, which, again, commute with the noise pattern. Therefore, the noise pattern on the bipartite state is applied to the final graph state without any modifications.

Second, we also need to take into account that the state $\Ket{G^\prime}$ itself might be noisy. Again, any Pauli-diagonal noise on that state can be written as correlated $Z$-noises. Therefore it suffices to considers what happens if qubit $3$ is affected by $Z$-noise. It holds that $\mathrm{CNOT}^{2 \rightarrow 3} Z^{(3)} = Z^{(2)} Z^{(3)} \mathrm{CNOT}^{2 \rightarrow 3}$. Therefore, $Z^{(3)}$ translates to $Z^{(2)}Z^{(3)}$ after the CNOT operation is applied. Again, the outcome of the $Z$-measurement and possible correction operations are not affected at all. So finally, every noise on the initial state $\Ket{G^\prime}$ that contained $Z^{(3)}$ translates to the same noise pattern with $Z^{(2)}$ on the final graph state.

Finally, applying these insights to our specific case gives rise to the noise pattern in \eqref{eqn:frombell_final_noise}.
 
\end{document}